\documentclass[manuscript,screen,nonacm]{acmart}

\AtBeginDocument{%
  \providecommand\BibTeX{{%
    \normalfont B\kern-0.5em{\scshape i\kern-0.25em b}\kern-0.8em\TeX}}}

\setcopyright{acmcopyright}
\copyrightyear{2018}
\acmYear{2018}
\acmDOI{XXXXXXX.XXXXXXX}

\acmJournal{JACM}
\acmVolume{37}
\acmNumber{4}
\acmArticle{111}
\acmMonth{8}

\acmPrice{15.00}
\acmISBN{978-1-4503-XXXX-X/18/06}

\begin{document}

\title{Test \& Evaluation Best Practices for Machine Learning-Enabled Systems}

\author{Jaganmohan Chandrasekaran}
\affiliation{%
  \institution{Virginia Tech National Security Institute}
  \city{Arlington}
  \country{USA}}
\email{jagan@vt.edu}

\author{Tyler Cody}
\affiliation{%
  \institution{Virginia Tech National Security Institute}
  \city{Arlington}
  \country{USA}}
\email{tcody@vt.edu}

\author{Nicola McCarthy}
\affiliation{%
  \institution{Virginia Tech National Security Institute}
  \city{Arlington}
  \country{USA}}
\email{nicmccarthy@vt.edu}

\author{Erin Lanus}
\affiliation{%
  \institution{Virginia Tech National Security Institute}
  \city{Arlington}
  \country{USA}}
\email{lanus@vt.edu}

\author{Laura Freeman}
\affiliation{%
  \institution{Virginia Tech National Security Institute}
  \city{Arlington}
  \country{USA}}
\email{laura.freeman@vt.edu}




\begin{abstract} Machine learning (ML) – based software systems are rapidly gaining adoption across various domains, making it increasingly essential to ensure they perform as intended. This report presents best practices for the Test and Evaluation (T\&E) of ML-enabled software systems across its lifecycle. We categorize the lifecycle of ML-enabled software systems into three stages: component, integration and deployment, and post-deployment. At the component level, the primary objective is to test and evaluate the ML model as a standalone component. Next, in the integration and deployment stage, the goal is to evaluate an integrated ML-enabled system consisting of both ML and non-ML components. Finally, once the ML-enabled software system is deployed and operationalized, the T\&E objective is to ensure the system performs as intended. Maintenance activities for ML-enabled software systems span the lifecycle and involve maintaining various assets of ML-enabled software systems. 

Given its unique characteristics, the T\&E of ML-enabled software systems is challenging. While significant research has been reported on T\&E at the component level, limited work is reported on T\&E in the remaining two stages. Furthermore, in many cases, there is a lack of systematic T\&E strategies throughout the ML-enabled system's lifecycle. This leads practitioners to resort to ad-hoc T\&E practices, which can undermine user confidence in the reliability of ML-enabled software systems. New systematic testing approaches, adequacy measurements, and metrics are required to address the T\&E challenges across all stages of the ML-enabled system lifecycle.

\end{abstract}



\maketitle

\section{Introduction}
ML has made significant strides in the past decade, leading to its adoption across various domains. ML includes methods spanning from classic statistical modeling approaches like linear models and decision trees to modern modeling approaches like deep learning. With ML, problems that were once too complex or impossible to handle and beyond the capabilities of traditional software systems can now be addressed. Thus, ML-enabled software systems are becoming increasingly prevalent. As more and more organizations across domains pivot towards leveraging ML-based solutions to address their needs, ensuring that ML-enabled software systems work as expected becomes increasingly important. 

\begin{figure*}
\centering
\includegraphics[width=0.7\linewidth]
{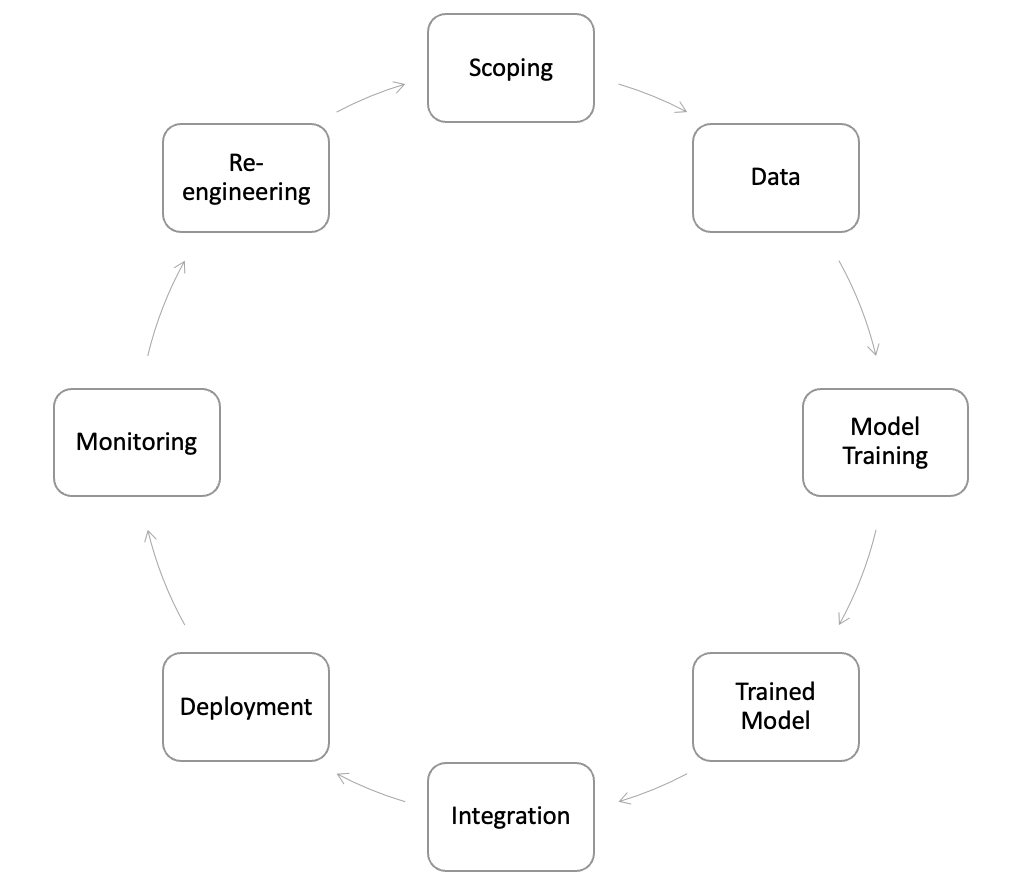}
\caption{Lifecycle of an ML-enabled software system}
\label{fig:lifecycle}
\end{figure*}

The lifecycle of an ML-enabled software system, as presented in Figure \ref{fig:lifecycle}, is a multi-stage process that involves scoping, data collection and processing, ML model development, integration, deployment, and post-deployment activities. The lifecycle commences with scoping, followed by data collection and processing. Once the data is prepared, practitioners begin the ML model development process. In this phase, the practitioner selects an ML algorithm from an ML framework, which is a collection of libraries and tools for building ML models. The ML algorithm, on receiving data and hyperparameters as inputs, analyzes and learns from the data in an iterative manner. The learned logic is referred to as the ML model. At the end of this phase, the practitioner has developed the ML component – a trained ML model. The ML model serves an input-output function in systems.

The next phase in the lifecycle involves integrating the ML component with other components of the software system. This integrated system is referred to as an ML-enabled software system. Subsequently, the ML-enabled software system is deployed in a production environment and utilized in inferencing activities. In the post-deployment phase, once the ML-enabled software system is operationalized, the system is constantly monitored, and maintenance activities, including re-engineering activities, are performed to ensure that the ML-enabled software system performs as intended and adheres to its desired operational standards. Developing an ML-enabled software system follows a cyclical lifecycle, and unlike traditional software systems, ML-enabled software systems may continue to evolve after deployment.

From a test and evaluation (T\&E) standpoint, there are fundamental differences between traditional software and ML-based systems. With traditional systems, a human turns requirements into decision logic using a programming language, and these systems exhibit a deterministic behavior. In contrast, the decision logic of ML systems is drawn from datasets by ML algorithms that analyze the underlying patterns in the data, often without human involvement. The decision logic, also referred to as an ML model, is essentially a black box, making it complex and difficult for humans to comprehend. Due to these factors, traditional T\&E approaches are not sufficient for ML-based systems.

\begin{figure*}
\centering
\includegraphics[width=1.0\linewidth]
{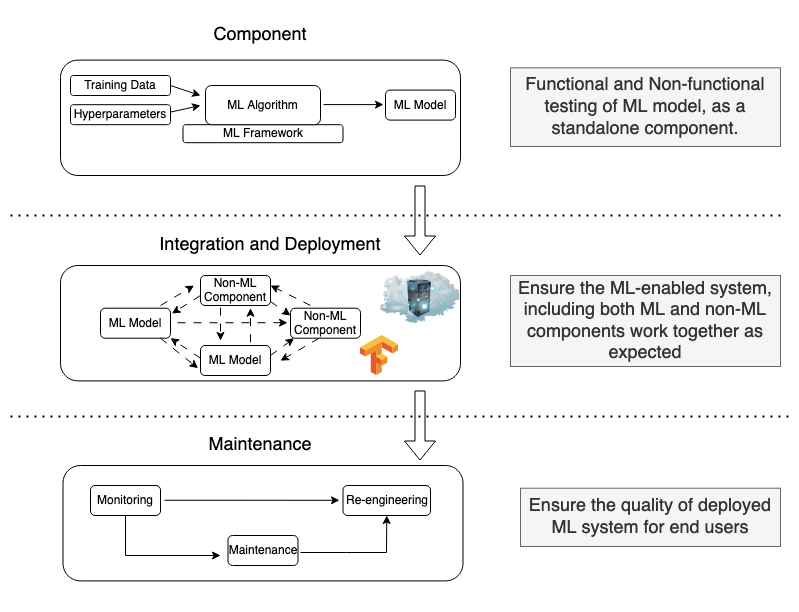}
\caption{Stages in T\&E  of an ML-enabled software system}
\label{fig:stages}
\end{figure*}

We review and discuss T\&E best practices for ML-enabled systems in this report. While the ML model is the core component of an ML-enabled software system, it is just one part of a larger software system. To harness the full potential of an ML-enabled software system, testing and validating the entire ML system across its lifecycle is essential. To this end, we take a holistic view of the challenges in testing and evaluating ML-enabled software systems. By drawing from the literature, this report provides a comprehensive exploration of testing challenges and strategies to address such challenges across the lifecycle of ML systems. As Figure \ref{fig:stages} denotes, we categorize the T\&E of ML-enabled systems into three levels: component level, integration and deployment, and maintenance \cite{CCMagazine}.

This report is structured as follows: Section \ref{Phase 1} presents the challenges and practices in the T\&E of the machine learning (ML) model at the component level. Section \ref{Phase 2} discusses the T\&E challenges associated with integrating and deploying ML-enabled software systems. Post-deployment T\&E challenges, including re-engineering activities, are covered in Section \ref{Phase 3}. Finally, Section \ref{conclusion} provides concluding remarks, and the Appendix contains a list of various tools used across the ML lifecycle.

\section {Component Level T\&E (Phase 1)} \label{Phase 1}

This section presents a summary of testing techniques that have been successfully adopted from traditional software testing to address the challenges in test generation for ML models. Furthermore, we examine the current practices in evaluating the performance of ML models on a test suite – metrics used for classifier and regression systems. Additionally, we explore the challenges of measuring the adequacy of a test suite, including the shortcomings of existing adequacy metrics. Finally, we present a set of metrics that have been specifically designed to measure the adequacy of test suites for ML-enabled systems.

A machine learning (ML) model is a key component of an ML-enabled software system. To build an ML model, the practitioner selects a learning algorithm and provides a training dataset and a set of hyperparameters as inputs. The learning algorithm, by analyzing the underlying patterns in the dataset, infers a decision logic iteratively \cite{chandrasekaran2021testing}. The learned decision logic is referred to as the ML model and serves as an input-output functional component. At the component level, the primary focus of T\&E activity is to evaluate the performance of the ML model itself.

Due to the fundamental differences between traditional software systems and ML-enabled software systems, T\&E for ML-enabled systems presents unique challenges. During the T\&E of ML-enabled systems, it is important to consider the following challenges.
\begin{itemize}
    \item \textbf{Data-Intensive Nature:} ML-enabled systems are data-dependent and, as a result, their performance relies heavily on the data used for the training and validation of ML models. Therefore, in addition to evaluating the decision logic of ML models, it is imperative to evaluate and ensure the quality of data employed throughout the ML lifecycle to ensure a high-quality ML-enabled system.
    \item \textbf{Large Input Space:} ML models derive their decision logic from the patterns underlying the training dataset. The general practice is to train ML models using a large training dataset, resulting in a large input space.  From a testing perspective, covering the large input space is challenging. Coverage of the input space requires a high number of test scenarios This results in a higher computing cost and effort, consequently making it infeasible to cover the input space.
    \item \textbf{Interpretable Decision Logic:} Unlike traditional software systems where a human writes the decision logic, the data-derived decision logic of ML systems is often not easily interpretable by humans. As a result, testers can find it difficult to effectively test and debug the behaviors of ML models.
    \item \textbf{Test Oracle:} One of the overarching challenges to testing ML systems is the problem of determining the expected outcome or ground truth. In most scenarios, test instances will lack a reliable test oracle and desired outcomes are unknown beforehand.
    \item \textbf{Non-Determinism:} ML systems exhibit stochastic behavior due to the statistical nature of their underlying learning algorithms. Their stochastic nature results in different outcomes for the same set of inputs over multiple runs. This non-deterministic behavior further exacerbates the challenges in effectively testing ML systems.
    \item \textbf{Emergent Behavior:} ML systems may learn and improve their behavior based on the input data they operate upon once deployed in real-world environments. While it may be a desired quality, the emergent behavior of ML systems introduces an additional set of testing challenges.
    \item \textbf{Beyond Correctness:} Given these unique characteristics, as discussed above, testing ML-enabled systems beyond correctness is necessary. There are many aspects regarding how these systems achieve outcomes that are relevant to T\&E. Some include:
    \begin{itemize}
        \item Fairness – the ability of the system to operate in an unbiased manner.  
        \item Safety – Given its emergent behavior, there is a need to guarantee that the ML system operates as intended and within the safe operating realm and does not cause harm.
        \item Explainability – Most ML systems derive their decision logic automatically, which is complex and not easily interpretable. From a T\&E standpoint, it is essential to guarantee that the decisions made by the model are understandable. 
        \item Security – As ML systems are data-dependent, providing privacy guarantees to data used across the lifecycle is paramount. 
        \item And others are meant to assure that the system operates in a reliable, robust, and trustworthy manner.
    \end{itemize}

\end{itemize}

Given these differences, traditional T\&E approaches may not directly apply to ML-enabled systems. Our survey of the literature shows that researchers and practitioners have adapted existing techniques for testing the component level. Given the scope of this document, we limit our discussion to current practices in test generation, test evaluation (metrics), and adequacy of the test suite used to evaluate ML models.

\subsection{Test generation:} Existing test generation techniques have been effectively applied to overcome the challenges in testing ML models \cite{zhang2020machine, riccio2020testing, ahuja2022testing}. Metamorphic testing, Differential Testing, Combinatorial Testing, and Fuzz Testing are some of the widely used traditional testing techniques that have been successfully adapted to test ML models, and they are discussed next. 

\textbf{Metamorphic testing:}  Metamorphic testing has become a widely used method to test ML systems. It relies on identifying metamorphic relations (MRs) based on domain knowledge or program specification. Researchers and practitioners have successfully adapted metamorphic testing-based approaches for testing ML systems \cite{segura2018metamorphic}.

To apply metamorphic testing, the MRs for the system under test are identified, and test cases that satisfy these relations are generated. The test cases are then executed and evaluated. In contrast to traditional testing techniques, where each test input is compared to the expected output, metamorphic testing evaluates test cases based on their ability to satisfy MRs. A test is deemed successful if it satisfies the metamorphic property; otherwise, it is classified as a failing test case. This approach eliminates the need for a test oracle and is particularly useful for testing ML systems.

Metamorphic testing relies on the relationship between input and output. In a sense, metamorphic relations substitute as test oracles for testing ML systems. Metamorphic testing enables us to test a system under test, in this case, an ML system’s ability to handle different variations of the same input through metamorphic properties. The underlying assumption is that different variations of the same input share similar properties, and hence the system under test, in this case, an ML system, should produce outputs that do not violate the identified metamorphic relations. 

Metamorphic testing has been successfully applied to test ML libraries \cite{ding2017validating} and trained ML models \cite{ding2017validating, tian2018deeptest, du2018deepcruiser, zhang2018deeproad}, and has been used to detect bugs in ML systems. It has also been leveraged to test a wide range of ML algorithms, including MartiRank \cite{murphy2008properties}, Support Vector Machines (SVM) \cite{murphy2008properties}, naïve Bayes classifiers \cite{al2017validation}, and k-nearest neighbor classifiers \cite{al2017validation}. Furthermore, they are applied to test deep neural network (DNN) models used in self-driving cars \cite{tian2018deeptest, zhang2018deeproad, ramanagopal2018failing}, image classifiers \cite{dwarakanath2018identifying}, and unsupervised learning algorithms \cite{xie2020mettle}. 
 
In traditional testing, the absence of failures in a test suite does not necessarily indicate the absence of bugs in the system under test. Similarly, a test suite satisfying all the metamorphic relations does not necessarily mean the ML system is free of bugs. The effectiveness of metamorphic testing is determined by the quality of MRs. Strong domain knowledge is critical for deriving high-quality MRs that are relevant and appropriate for the system under test. 

Therefore, selecting appropriate MRs that are relevant to the system under test is crucial for the effectiveness of metamorphic testing. A thorough understanding of the system and strong domain knowledge are necessary for deriving high-quality MRs.

\textbf{Differential Testing:} Differential testing is a technique that aims to detect faults in software systems by comparing the performance of test cases on similar versions of software implementations. Differential testing compares different versions of software implementations \cite{mckeeman1998differential}. The fundamental idea is that given an input, similar implementations of the software system must produce an identical or similar output. If one of the implementations produces a different output, then the program is considered error-prone and needs further investigation for faults \cite{mckeeman1998differential}. Differential testing alleviates the need for a test oracle by comparing the outputs of similar implementations and treating the inconsistency among implementations as indications of faulty behavior.

Pei et al. leveraged differential testing to address the Oracle challenges in testing ML systems \cite{pei2017deepxplore}. Multiple ML models with identical functionalities (steering control in autonomous vehicle systems) are used as cross-referencing oracles. In their approach, given a test input, its output is compared across multiple ML models. If the output of one of the ML models differs from the rest, then the test input is considered failure-inducing \cite{pei2017deepxplore}. Similarly, differential testing has been used to test and evaluate automatic speech recognition systems \cite{asyrofi2020crossasr}. Pham et al. proposed CRADLE, which leverages differential testing to test and evaluate ML frameworks \cite{pham2019cradle}. CRADLE was evaluated on three deep learning frameworks, TensorFlow, CNTK, and Theano, using 11 datasets and 30 ML models. The authors identified multiple bugs and numerous inconsistencies (>100) among the three frameworks. Herbold et al. \cite{herbold2023differential} used a differential testing-based approach to T\&E non-deep learning classification algorithms implemented on four widely used ML libraries, namely scikit-learn \cite{scikit-learn,sklearn_api} WEKA \cite{hall2009weka,eibe2016weka}, SparkMLLib \cite{mllib,meng2016mllib}, and Caret \cite{caret}. 

Studies from the literature suggest that differential testing has been successfully leveraged to test the correctness of a trained ML model \cite{pei2017deepxplore, asyrofi2020crossasr, guo2018dlfuzz} and the correctness of ML frameworks or libraries used for training ML models \cite{pham2019cradle, herbold2023differential}.  

While differential testing-based approaches ease the challenges of test oracle, they can only be applied if there is more than one implementation of the same functionality. Furthermore, even when multiple implementations are available, a fault is only detected if at least one of those implementations results in a different output compared to the others

\textbf{Combinatorial testing:}  In an ideal scenario, the test and evaluation of a software system involves assessing the system’s performance under all possible conditions, also referred to as exhaustive testing. However, in most cases, exhaustive testing is neither feasible nor practical, given the cost and time limitations. Combinatorial testing is a technique that aims to systematically test all possible interactions of input parameters of the system under test (SUT). In other words, combinatorial testing enables testers to perform pseudo-exhaustive testing.  

Given a software system with “t” input parameters, the combinatorial testing technique generates a test set in a manner such that it guarantees that each “t-way” combination of input parameters is covered at least once by the test set. Given a system under test, combinatorial testing enables practitioners to test various combinations of its input parameters systematically. Furthermore, combinatorial testing can significantly reduce the number of test cases generated while maintaining its defect detection capability. Empirical studies have demonstrated that most failures in software systems can be attributed to the interactions between relatively fewer input parameters \cite{kuhn2010practical}. That is, in most cases, interactions between certain input parameter values trigger failures, referred to as interaction failures. Using combinatorial testing techniques, practitioners can detect these interaction failures in software systems, and existing studies suggest that interactions among six or fewer features are sufficient to trigger most failures in software systems \cite{kuhn2004software, kuhn2010practical}. 

Combinatorial testing has been adapted to address the test \& evaluation challenges in ML-enabled systems.  Combinatorial testing has been leveraged in assessing the quality and comparing datasets \cite{tuncali2018simulation}, slicing datasets \cite{ackerman2020freaai,barash2019bridging}, test set design \cite{cody2022systematic}, synthetic data generation \cite{khadka2023synthetic}, testing implementations of traditional ML algorithms \cite{chandrasekaran2017applying, herbold2022smoke}, testing pre-trained DNN models \cite{tuncali2018simulation,ma2019deepct,chen2019variable, chandrasekaran2021combinatorialADS,gladisch2020leveraging,li2020ontology}, active learning \cite{katragadda2022active}, transfer learning \cite{ahamed2021attl}, explaining decisions of AI models \cite{kuhn2020combinatorial, chandrasekaran2021combinatorial}, evaluating fairness behavior of ML models \cite{patel2022combinatorial}.

Given its ability to systematically test large-scale software systems with a relatively lesser number of test cases, combinatorial testing can be leveraged to address the challenges caused by the large input space of the ML-enabled systems. However, the key to the effectiveness of combinatorial testing in testing ML-enabled systems relies on the modeling of the input parameters and their corresponding values during the test generation process

\textbf{Fuzz Testing:} Fuzz testing aims to detect program crashes, memory corruption, and security issues in software systems.  The purpose of “fuzzing” is to identify potential vulnerabilities and assess the robustness of a software system by generating random and unexpected inputs for the system under test. By subjecting the software system to diverse and unexpected inputs, fuzzing can uncover weaknesses in software systems that would otherwise go undetected by traditional testing techniques.

Fuzzing has been extended to test ML-enabled systems. Guo et al. propose a fuzzing-based test generation tool that leverages coverage feedback from the DNN architecture to generate adversarial inputs to test DL systems \cite{guo2018dlfuzz}. Similarly, Xie et al. propose a test framework that leverages coverage-guided fuzzing to generate inputs to test and evaluate DNNs \cite{xie2019deephunter}, and Odena et al. introduce an approach that combines coverage-guided fuzzing to debug DNN models \cite{odena2019tensorfuzz}.  Zhou et al. propose a test generation strategy that integrates metamorphic testing and fuzzing and successfully detected critical errors in an obstacle perception module used in autonomous driving systems \cite{zhou2019metamorphic}.  Fuzzing techniques have also been used to evaluate the implementations of deep learning libraries \cite{yang2023fuzzing,xie2022docter,deng2022fuzzing,wei2022free}.

In general, fuzzing can be leveraged to test ML-enabled systems in the following ways: 
\begin{itemize}
    \item To evaluate ML-enabled systems on random out-of-distribution inputs that might not be covered by a traditional test suite (outlier scenarios)
    \item To identify potential security vulnerabilities in ML-enabled systems, and 
    \item To thoroughly evaluate the robustness of ML-enabled systems.
\end{itemize}

\textbf{Adversarial Testing:} ML-enabled systems are prone to adversarial attacks in which a malicious actor can provide an input identical to a natural input yet induce the ML system to exhibit an undesired behavior. Adversarial inputs tend to consist of subtle changes compared to natural inputs, which can be harder to detect due to their subtleness. Studies from the literature have demonstrated that ML-enabled systems, including state-of-the-art ML models, are susceptible to adversarial attacks. Szegedy et al. demonstrated that it is possible to manipulate the behavior of an image classifier by providing adversarial samples that are indistinguishable from a natural input sample \cite{szegedy2013intriguing}. 

From an operational standpoint, it is essential to validate that the ML system, particularly the ML components, is resilient to adversarial attacks. Hence, in addition to correctness, ML-enabled systems are evaluated for their robustness against adversarial inputs. To this end, testers generate adversarial inputs and evaluate the ML system’s robustness against adversarial attacks. Adversarial test inputs are generated by applying undetectable perturbations to natural inputs. Existing techniques, such as Fuzz and Differential testing, have been adapted to generate adversarial test inputs \cite {guo2018dlfuzz, zhou2020deepbillboard, tuncali2018simulation}. Furthermore, Generative Adversarial Network (GAN) was combined with a metamorphic testing technique to generate adversarial test inputs \cite{zhang2018deeproad}. Cleverhans, ART, DeepRobust , FoolBox, SecML, AdverTorch,  are a few of the open-source libraries for generating adversarial instances and evaluating the robustness of ML-enabled software systems \cite{croce2020robustbench}.

To summarize, traditional testing techniques successfully address the test generation challenges in ML-enabled systems. Specifically, metamorphic testing and differential testing aim to address the test oracle problem, while combinatorial testing aims to address the large input space problem. In some cases, a combination of these techniques such as Metamorphic Testing+Fuzzing, Combinatorial+Fuzzing, and Fuzzing + property-based testing has been proposed to address the testing challenges. In addition to testing the model implementation, a significant amount of work has been reported on testing the ML libraries \cite{zhang2020machine}.

\subsection{Metrics:} The primary goal of any testing activity is to determine if the system under test performs as expected. One of the primary objectives of T\&E of ML-enabled systems is to evaluate and determine the readiness of ML-enabled systems for usage in the real world.  In a sense, T\&E of ML-enabled systems seeks answers to questions such as 1) Is the ML model good enough? 2) Can the ML system exhibit a reliable performance? 

Given the statistical nature of ML systems, existing metrics or criteria used to evaluate the performance of traditional software systems might not suffice for ML-enabled systems. Specifically, a common expectation in T\&E of traditional software systems is that the SUT is considered good to integrate or used in the real world if ALL the test cases are passed. This is not the case in ML-enabled systems. As they are statistical in nature, the sufficiency criteria for these systems are different in the sense that an ML-enabled system is considered to be good enough to be deployed or used in the real world as long the overall performance of the system on a test suite is above a satisfactory threshold, which the practitioner or engineer defines. This is different from traditional software systems. In the case of the traditional software systems, each test case from the test suite must pass for the SUT to be considered good enough, while in the case of the ML-enabled systems, it is not necessary that all test cases must result in a passing state. Instead, the overall performance of the test suite is considered for evaluating the performance of the SUT.

Practitioners use a set of metrics to evaluate the performance of an ML model on a test set. The most commonly used metrics to evaluate classifier systems include Accuracy, Precision, Recall, and F1 score. Meanwhile, regression-based ML systems are evaluated using Root Mean Square Error (RMSE), Mean Square Error (MSE), and Mean Absolute Error (MAE) \cite{hutchinson2022evaluation, zhengxin2023mlops}.

\textbf{Limitations of current evaluation metrics:} It is to be noted that the testing objective of an ML engineer and a Test engineer differs. ML engineers are focused on maximizing the ML model performance on representative datasets, while test engineers aim to test the ML system comprehensively and to ensure that the system behaves as expected and meets user requirements. However, in practice, both ML engineers and test engineers use similar or identical metrics to evaluate the performance of ML systems. Although the current practice helps evaluate the models for T\&E purposes, the current set of metrics may not truly or adequately evaluate model capabilities according to the domain or environment in which it will be operated. 

Furthermore, evaluating ML systems developed for different domains and use cases with the same set of metrics might not be the right approach. Over-reliance on a set of metrics might adversely impact the T\&E process as they might fail to assess the ML system readiness truly. One key question to consider is whether the current state-of-the-art metrics effectively evaluate the capabilities of ML model relevant to the real-world environment in which it will be operated upon? As suggested by Hutchison et al. metrics should be grouped into learner-focused or application-centric metrics. The former group primarily focuses on assessing the learning ability of the ML model, while the latter evaluates the ML system’s ability to operate safely and reliably in the environment in which it will be operated upon \cite{hutchinson2022evaluation}.

In conclusion, there are established metrics available to evaluate the learning capabilities of ML models. However, to the best of our knowledge, there is currently a lack of application-centric metrics. Further research is necessary to develop metrics that can accurately assess the operational safety and reliability of ML-enabled systems in real-world environments.

\subsection{Test adequacy:}  With the availability of numerous test-generation techniques, practitioners may face a challenge in creating or selecting an efficient and comprehensive test suite that enables them to evaluate the overall quality and reliability of the software.

Test adequacy is a measure to evaluate the rigor and thoroughness of the testing effort. Given a test suite, the primary goal of the test adequacy metric is to enable the practitioner (in this case, the tester) to ascertain whether the test suite has sufficiently tested the software under test and to identify any potential gaps that may exist in the testing process. In a sense, test adequacy helps to evaluate if a test suite is sufficient.  Higher test adequacy suggests the test suite is sufficient, while a lower value suggests the system is not adequately tested enough.

In the case of traditional software systems, structural coverage metrics such as line coverage, statement coverage, branch coverage, and MC/DC coverage are widely adopted as test adequacy criteria. For example, MC/DC coverage is used as test adequacy criteria for testing mission-critical systems. Given the fundamental differences between traditional software systems and ML-enabled systems – decision logic not being explicitly encoded and represented in the form of programming statements, existing test adequacy, which are structural-based metrics, will not be applicable to ML-enabled systems. 

Recall that ML-enabled systems are data-intensive systems; data and ML algorithms are the two key components of an ML-enabled system. Researchers have proposed adequacy criteria that consider these characteristics of ML-enabled systems, and they are discussed next.

\textbf{Neuron Coverage:} Similar to structural coverage criteria like branch or line coverage in traditional systems, Pei et al. proposed neuron coverage as a test adequacy criterion for DNN systems \cite{pei2017deepxplore}. Neuron coverage measures the number of neurons activated to the total number of neurons in the DNN model \cite{pei2017deepxplore}. The underlying belief is that, as interactions between neurons across layers results in model output, if a test suite activates every or a sufficient number of neurons, then it is considered to execute all the possible executed path and thus can be treated as equivalent to structural coverage metrics for traditional software systems. Furthermore, Ma et al. proposed a set of multi-granularity coverage criteria, namely k-multisection neuron coverage, neuron boundary coverage, strong neuron activation coverage, Top-k Neuron coverage and Top-k Neuron Patterns to assess the quality of the test set and guide in the test generation process \cite{ma2018deepgauge}. Later studies demonstrated limited or no correlation between neuron coverage and test adequacy \cite{li2019structural, harel2020neuron}. A test suite achieving a higher neuron coverage might not necessarily test the DNN model sufficiently. Using neuron coverage as test adequacy and generating test cases does not improve the quality of the test set. In some cases, test cases generated based on neuron coverage are not natural or realistic, given the operating domain.

\textbf{Surprise Adequacy:}  To address the limitations of structural coverage-based adequacy metrics, Kim et al. proposed a new metric referred to as surprise adequacy \cite{kim2019guiding}. This metric measures test adequacy based on the diversity of the test set. The fundamental idea behind this metric is that measuring the adequacy of a test set must be based on the surprising nature of the test inputs compared to the model training set. A surprising test input should be different from the training instances that the model has previously seen but not too different from the distribution of the training data. Furthermore, a good test suite must encompass instances similar to the training set and previously unseen, surprising inputs. Evaluating the model on diverse inputs helps assess behavior on out-of-distribution inputs. The surprising adequacy of input is measured based on two metrics: Likelihood-based Surprise Adequacy (LSA) and Distance-based surprise adequacy (DSA). Both metrics aim to measure the novelty of test inputs relative to the dataset the model was trained in. 

Although the surprise adequacy metric is not purely structural, meaning it does not assess the test adequacy purely based on the structural coverage of DNNs, the efficacy of this metric depends on the underlying DNN architecture. Additionally, the metric is computationally expensive and may not scale for large datasets, rendering the metric impractical for ML-enabled systems with larger training datasets. 

\textbf{Combinatorial Coverage:}  Data is a key component of an ML system. Therefore, it is prudent to assess the adequacy of a test set based on the data rather than solely relying on structural coverage. Lanus et al. proposed combinatorial coverage (CC), a metric that evaluates dataset sufficiency based on the presence or absence of feature interactions in a dataset compared to a user-defined universe. Additionally, they proposed an extension of combinatorial coverage, referred to as set difference combinatorial coverage (SDCC) that allows a practitioner to identify similarities and differences between two given datasets in the same universe \cite{lanus2021combinatorial}.

From a T\&E perspective, the adequacy metric should assess two important aspects. Firstly, it should evaluate the degree of similarity or dissimilarity between the test set and the training set. If a test set closely resembles the training set, it fails to effectively evaluate the model’s capabilities in unfamiliar conditions that it is likely to encounter once deployed. However, on the other hand, the test set should not deviate significantly from the training set.  Secondly, the adequacy metric should evaluate the alignment between a test dataset and the operational environment. This involves measuring how closely the test dataset mirrors the conditions and characteristics of the actual operational environment.

Current metrics, such as SA and SDCC, assess the adequacy of a test set by examining the relationship between the training data and test instances. However, these approach does not take into account the operational environment in which the model will be utilized, which is a critical factor in testing and evaluation (T\&E) from an operational perspective. This limitation is a significant drawback of existing adequacy metrics, which fail to evaluate the adequacy of a test set with respect to the operational environment of the model.

To summarize, practitioners have successfully adapted existing testing methods to test ML models. However, generating valid and meaningful test cases remains a challenging task. Unlike traditional systems, test data in ML-enabled systems may be repurposed as synthetic training data for future training activities, necessitating the generation of realistic and valid test data. Given the nature of ML-enabled systems, evaluating the system beyond functional correctness is necessary. Furthermore, test adequacy for ML-based systems is still in its infancy, and there is a need for new approaches to “truly” assess the adequacy of test suites. Additionally, current metrics to evaluate ML model performance may fall short of evaluating the model’s learning capabilities. Rather, they may evaluate the model’s memorization capabilities. Therefore, new metrics are necessary to evaluate the learning capabilities of ML models.

\section {Integration and Deployment (Phase 2)} \label{Phase 2}

This section explores the complexities and T\&E challenges associated with integrating an ML model with other software components, as well as deploying the resulting ML-enabled software system. We present an overview of two widely used techniques and outline the T\&E challenges faced by the practitioners while using the techniques. Furthermore, this section briefly discusses the common failures encountered in this phase and the gaps in current practice.

The deployment phase in the ML lifecycle involves integrating one or more trained ML models with other software components and deploying the ML-enabled system. Several challenges that can emerge in this phase include seamless integration of ML and non-ML components, managing platform dependencies, scalability, and reliability. Practitioners follow two common techniques to integrate and deploy ML models: model conversion and model quantization. Model conversion techniques enable interoperability within the ML ecosystem. It involves converting an ML model from one format to another for cross-platform deployment. On the other hand, model quantization techniques are used to optimize or reduce an ML model's size (memory footprint) and computation requirements for deploying in resource-constrained environments \cite{chen2020comprehensive}.

Deploying ML-enabled systems is considered a challenging task \cite{chen2020comprehensive,baier2019challenges}. Depending on the field of operation, an ML-enabled system may be deployed and accessed through different platforms, such as a cloud server, on a mobile device, or through a browser \cite{chen2020comprehensive}. Cloud servers provide sufficient hardware and scalable infrastructure capabilities for ML-enabled system deployment among the platforms. Application Programming Interfaces (APIs) enable inferencing on ML models deployed on cloud servers. As a result, cloud servers are the preferred deployment platform among practitioners \cite{chen2020comprehensive, zhang2019software}. Unlike cloud servers, deployment platforms, such as mobile devices, are resource-constrained, and executing ML models (or DNN models) in their original size will not be feasible. To overcome these limitations of resource-constrained operational environments, pre-trained ML models are compressed (quantized) before deployment. 

CoreML \cite{CoreML} and TF Lite \cite{TFLite} are widely used frameworks to facilitate the model deployment process on mobile devices. Using model compression may introduce performance issues in ML models \cite{guo2019empirical}. Practitioners observe a difference in outputs between the two versions – the original model and the compressed ML model. Quantization techniques, which reduce the precision of model weight representations, may cause a compressed ML model to have a variation in the output compared to its original counterpart. Therefore, it should be noted that all mismatches between the original and converted version shall not be considered as the presence of faults \cite{chen2021empirical}.  

Furthermore, in certain scenarios, faults introduced through the quantization process do not necessarily trigger a failure or impact the inference capabilities of the ML-enabled system. Rather, practitioners may observe a drop in performance in the form of degradation in inferencing speed or excessive resource (memory) utilization \cite{guo2019empirical, chen2021empirical}. Thus, one of the common T\&E challenges is validating the performance of compressed models, specifically evaluating the differences in performance between the original ML model and the converted ML model. 

Hu et al. studied the performance degradation in quantized ML models and analyzed the data characteristics that induce the performance differences between the original and quantized models \cite{hu2023towards}. They studied the behavior of the compressed ML models on data distribution shifts, a phenomenon experienced by the ML models once operationalized.  Furthermore, they studied the effect of different training strategies on the behavior of the ML model and its respective compressed counterparts. They evaluate the performance of both the original and compressed versions by measuring the prediction disagreements between them. Results from their study suggest that models trained using the quantization-aware training approach exhibit better performance than those using other training approaches.

In the case of a compressed model exhibiting poor performance, one of the widely followed approaches is to retrain the model. However, retraining a quantized model might not be feasible. This can be explained as follows: Quantization reduces an ML model's parameters and internal configurations, thus reducing its memory requirements. As a result, retraining the quantized model might result in suboptimal performance. Hence, among the debugging strategies currently followed by the practitioners, a common go-to strategy is to repair the original ML model even though an issue might have been discovered/identified during/due to model conversion, integration, or inference after integration \cite{chen2021empirical}. 

Next, we discuss the challenges the practitioners face while deploying ML models across different platform configurations. Version Compatibility within the ML ecosystem is one of the major hurdles practitioners faces while deploying ML systems \cite{guo2019empirical}. With multiple ML frameworks available for ML model development, in most cases, ML models may be trained using an ML framework and deployed on a different ML framework. The general practice is to leverage model conversion tools to convert the pre-trained ML model into a format supported by the deployment platforms. TensorFlowLite \cite{TFLite}, Open Neural Network eXchange (ONNX) runtime\cite{ONNX}, Caffe2 \cite{caffe2}, NVIDIA TensorRT \cite{Nvidia}, MXNET Model server \cite{mxnet} are a few of the widely used frameworks or tools to perform the model conversion.  For example, practitioners use model conversion tools to deploy ML models on cloud servers \cite{chen2020comprehensive}.

While the model conversion tools aim to facilitate the deployment and integration activities in the ML lifecycle, they may inadvertently introduce errors that could potentially affect the performance of ML models \cite{zhang2019software, guo2019empirical}. For example, differences in computations resulting in prediction errors may arise due to variations among ML frameworks in processing and handling internal operations in ML models. 

Chen et al. analyzed model deployment faults discussed in software developer forums, namely Stack Overflow and GitHub \cite{chen2021empirical}. Of the 304 faults analyzed, 185 faults occur during model conversion or model integration activity, with around 48\% of total faults reported in the model conversion activity. Furthermore, the authors grouped the model conversion errors into twelve categories, with the unsupported operation being the most frequent failure-triggering operation in the model conversion process. During the model integration phase, framework loading failure, model loading failure, GPU delegate failure, or Dependency resolution error are a few of the commonly reported failures, with dependency resolution error accounting for the most failures in this phase.

Furthermore, version compatibility issues can trigger runtime errors, crashes, or memory issues.  A study conducted by Jajal et al. analyzes the failures of the ONNX ML model converter \cite{jajal2023analysis}. The findings suggest that models converted using ONNX are generally stable enough for usage; it is imperative to subject the converted ML models to rigorous testing \& evaluation before their deployment in the real world. Furthermore, these findings underscore a gap in current practices, highlighting the need for future research to develop methods and approaches to test and evaluate ML models during deployment effectively.

\textbf{Data Mismatch:} After model deployment, one of the challenges often faced by practitioners is handling errors that arise due to the mismatch in input data.  This can be explained as follows: Testing in the model development phase focuses on evaluating the model with test inputs conforming to pre-defined data formats. However, in real-world scenarios (deployed environments), the ML-enabled system may encounter inputs that may not conform with the anticipated input formats, and such input data mismatch may trigger a failure or unexpected behavior of the ML system \cite{chen2020comprehensive, guo2019empirical, openja2022empirical}. To mitigate the risks in handling inconsistent data formats, practitioners can incorporate data pre-processing wrappers or scripts into the deployment pipeline. This shall ensure incoming data undergoes requisite transformations, aligning it with the model's expected format.

\textbf{Security Challenges in Model Deployment:} Deploying ML models in unencrypted form makes them susceptible to security attacks. For example, an adversary can launch a model extraction attack to steal information about the internal of the ML model or to launch a model poisoning attack. To mitigate security risks in the ML pipeline, one of the widely adopted approaches is differential privacy, a mathematical framework to safeguard data privacy \cite{dwork2006our, dwork2008differential}. Using a differential privacy framework, practitioners introduce noise to model gradients, loss function, or model weights during the training phase or train the ML model and introduce noise to its inference \cite{zhu2020more}. 

\textbf{Quality guarantees do not translate:} The quality guarantees of ML models ascertained at the component level do not necessarily translate to the converted models \cite{baier2019challenges, guo2019empirical, hu2023towards}.  Compared to the original model, its quantized version may exhibit a reduction in prediction performance (drop-in accuracy), increased resource utilization, longer time to perform inference, and a drop in robustness \cite{chen2021empirical}. 

\textbf{Deployment-specific testing:} At the component level, the focus of T\&E is to evaluate the functional correctness of ML models. However, once deployed in a real-world setting, the ML system shall be exposed to outliers and surprising input scenarios. The ML system must be thoroughly evaluated for its ability to detect and handle outlier events. The robustness of ML systems is a major focus of interest from a T\&E perspective. 

\textbf{Need for realistic test evaluations and application specific metrics:} 
In general, practitioners may use traditional ML metrics across various domains to evaluate the performance of models. However, this approach can be problematic and one of the common pitfalls in ascertaining the readiness of ML models for deployment. The issue lies in the fact that these traditional metrics can be misleading and may not “truly” capture or evaluate the model’s ability to perform tasks once deployed in the real world. Instead, these metrics are primarily designed to assess the learning capabilities of the model. This creates a gap in the implicit assumption that if a model can learn, it can automatically transfer the knowledge to perform a task. The current practice of using traditional ML metrics to evaluate model readiness does not fully evaluate this assumption. Therefore, new testing methods and evaluation metrics are needed to evaluate this assumption systematically and to successfully deploy ML-enabled systems in the real world. 

To sum up, model conversion and model quantization are the main techniques practitioners leverage to integrate and deploy ML-enabled systems. In this phase, common challenges practitioners encounter include behavior mismatch between the original model and converted or compressed model, non-failure performance degradation such as higher resource utilization or longer inferencing time, program crashes and restarts (unstable software), and security issues. Although there might be variations in ML performance due to conversion and compression frameworks, not all variations shall be treated as the presence of bugs. Thus, performing T\&E in this phase is a significant challenge.

To the best of our knowledge, compared to the T\&E of the ML model at the component level, only a limited amount of work is reported on the T\&E of the ML model at the integration and deployment stage. Future research is needed for further exploration and development of systematic approaches to address the challenges related to the testing and evaluation of ML-enabled systems during model deployment. 

Additionally, the current T\&E literature primarily focuses on understanding the impact of prediction performance in quantized models. In the future, it will be worth examining how model quantization and model transformation activities impact the assurance characteristics, such as Explainability, fairness, and security in ML systems.  We need new methods and metrics for T\&E at the deployment stage. Also, additional methods and metrics are needed to evaluate the ML model’s ability to handle Out-of-Distribution data, unexpected input data format, and inferencing time. 

\section {Maintenance (Phase 3)} \label{Phase 3}

This section delves into the T\&E challenges that arise during the maintenance phase of ML-enabled software systems. Given the characteristics of ML systems – emergent and data-dependent – it is essential to ensure that the ML-enabled system continues to perform as intended post-deployment. We outline the maintenance challenges that are unique to ML-enabled systems and discuss the current practices and challenges in managing various assets of ML-enabled systems. Furthermore, we discuss the current re-engineering practices adopted by practitioners and present the T\&E challenges encountered by practitioners. We conclude by identifying the gaps in the current practices in the ML-enabled system maintenance activities.

Once the ML-enabled system is deployed in the real world, it is critical to ensure consistent performance over time. In this phase, an undesired behavior can arise due to multiple factors. Firstly, buggy implementations of ML models can impact the behavior of the ML systems. Secondly, given the emergent behavior of ML systems, the input data may adversely impact the ML model’s behavior. Lastly, deviations in the statistical properties of the data used for training the ML model and the input data encountered in the operational environment may contribute to the undesired behavior of ML systems. From a T\&E perspective, the primary objective is to ensure the deployed ML-enabled systems continue to perform as expected. To this end, in this phase, practitioners perform maintenance activities to mitigate the aforementioned challenges and ensure the reliable performance of ML systems in the deployed environment.

Next, we present the maintenance challenges in ML-enabled software systems. Traditionally, it is estimated that maintenance activities in software systems consume a significant amount of the project cost, time, and effort \cite{shivashankar2022maintainability, granja1997method}. In addition to the traditional software maintenance challenges, characteristics of ML-enabled systems, such as data-dependent, emergent behavior, and decision logic represented in a complex form, add another layer of maintenance challenges that are unique to ML-enabled systems. For example, in a traditional system, maintenance activities shall begin once the software system is fully developed, tested, and deployed. However, as \cite{shivashankar2022maintainability} suggests, maintenance activities in ML systems transpire across all stages of their lifecycle, from the data phase to the operational phase, and are discussed next. 

Data plays a critical role in the performance of ML systems. In most cases (if not all), data in its original format may not be suitable for building ML systems. Using a data preprocessing pipeline, practitioners transform the original data into a format suitable for building ML systems. In the development phase, the practice is to train and fine-tune the model’s performance iteratively. So, in the case of a model exhibiting poor performance, a common strategy employed by the practitioners is to reevaluate the choice of data features and/or hyperparameters used in the training process. This necessitates capturing and cataloging all the intermediate steps in the training process, including data preprocessing, choice of hyper-parameters, and ML algorithms to ensure reproducibility. To this end, data and model versioning tools can be leveraged to systematically track different versions of data and ML models to facilitate reproducibility.

\textbf{Test Data Maintenance in ML-enabled systems:} As models are developed and deployed frequently, it is necessary to ensure that the test data is up-to-date and reflective of the operational environment. As a part of maintenance activity, test data must be updated periodically. Failure to do so might result in evaluating the ML models using a stale dataset and consequently, resulting in performance issues that might arise from using a stale dataset in the testing process.  Furthermore, with ML-enabled systems constantly evolving in each iteration, test data used across the pipeline, such as test data in the training phase, test data used in the deployment and monitoring phase, regression testing data as well as the metrics used in evaluating the performance, must be updated to accurately reflect the environment in which the model will be operated upon.

\textbf{Glue code:} Across the various stages in the ML pipeline, practitioners use different software frameworks/libraries to accomplish tasks. Practitioners use ad-hoc programs, referred to as glue code, to coordinate and manage the interactions among different components and facilitate a seamless integration and workflow across the ML pipeline. As the name suggests, glue code connects all different components together in the ML pipeline. Currently, there are no specific standards for developing and maintaining the glue code. Since glue code is developed in an ad-hoc manner based on the chosen framework or library, it tends to be a prominent source of technical debt in the ML pipeline \cite{shivashankar2022maintainability}. Maintaining the glue code is another challenge unique to ML systems, as bugs could stem from the glue code. To the best of our knowledge, no existing research has explored the T\&E challenges that may arise from the use of ad-hoc glue code.

\textbf{Current gaps in ML Maintenance:} Given that data significantly influence the behavior of ML systems, first as a training dataset during the modeling process and later once deployed via a feedback loop, it is prudent to approach the maintenance practices from a data-first approach. Furthermore, as maintenance activities span across the ML pipeline, maintaining various assets of an ML system – data, choice of hyperparameters, ML algorithm, ML framework configurations, deployment configurations, test data, and metrics is a challenge faced by practitioners. There is a need to develop approaches and tools to streamline and standardize the maintenance activities across the ML pipeline.

\textbf{Performing robustness checks of ML systems in the maintenance phase:} Unlike traditional systems, the performance of an ML-enabled system can shift after deployment due to its emergent behavior.  Performing periodic robustness assessments during the maintenance phase can help ensure the deployed system continues to perform as expected. Chen et al. proposed a conceptual framework called AI Model Inspector to inspect the robustness of AI systems \cite{chen2023ai}. The framework draws parallels from the automotive industry and is divided into four phases analogous to the stages in the development and maintenance of automobiles. In addition to the traditional development and maintenance activities, the authors advocate for educating users on AI, such as ethics and value alignment, before using AI systems, similar to obtaining a driver's license before operating an automobile on roads.

The proposed framework evaluates the robustness of AI models in two phases: Detection and Mitigation. The framework runs routine maintenance checks on the AI model in the detection stage to identify potential issues. This is similar to periodic maintenance checks performed on automobiles. Next, in the mitigation phase, if necessary, the framework applies the necessary model fixes, including retuning or retraining models and reevaluating the AI model’s robustness. Note that, although the author proposes the conceptual framework for AI systems in general, given the scope of this manuscript, the suggestions presented in this manuscript are limited to ML-enabled systems.

Furthermore, deviating from the customary norm of evaluating a software system through a binary lens – whether the system is deemed either robust or non-robust -- the authors draw inspiration from the different stages of autonomy displayed by autonomous vehicles. They propose six levels to evaluate the robustness of ML-enabled systems, ranging from no robustness (Level 0) to human-aligned and augmented robustness (Level 5). 

Under current practices, addressing performance issues in deployed ML systems, including fixing bugs, is reactive in nature. This means practitioners typically engage in analyzing and identifying necessary mitigation efforts when an issue has been identified or reported. However, given the broader adaption of ML-enabled systems across domains, including safety-critical domains, postponing maintenance issues until issues manifest and relying on post-incident mitigation efforts may prove inadequate. Instead, developing a proactive maintenance framework for ML-enabled systems is essential, like the conceptual framework presented by \cite{chen2023ai}. Such proactive frameworks can periodically evaluate the ML systems, monitor the model’s health, reveal potential vulnerabilities, and proactively address identified issues through corrective actions before they impact the performance of ML-enabled systems.

\textbf{Monitoring:}
Given its unique characteristics, such as data-driven and emergent behavior, ML-enabled systems may experience changes in their behavior once deployed, necessitating post-deployment activities to provide performance guarantees. Monitoring the behavior of the ML-enabled system, specifically ML models, is essential to detect performance degradation, biased behavior, or noncompliance with system regulatory requirements \cite {zhengxin2023mlops, giray2021software, wan2019does, nahar2023meta}. Additionally, monitoring helps identify excessive resource or infrastructure utilization.

Upon deployment, the machine learning (ML)-enabled system may be subjected to varied data patterns in the operational environment, which deviate from the data it was initially trained with. This phenomenon is referred to as drift in ML-enabled systems, and they can be broadly categorized as data drift and concept drift \cite{zhengxin2023mlops}. Data drift occurs when the statistical properties of training data change, while concept drift occurs when the relationship between "features" and the target label changes over time. Concept drift is often more subtle than data drift, making it harder to detect. Furthermore, concept drift is susceptible to adversarial exploitation, where a malicious actor may deliberately introduce a concept drift to manipulate the behavior of the ML-enabled system \cite{nelson2015evaluating}.  

ML-enabled systems can experience an abrupt change in distribution, referred to as sudden drift, or the distribution changes progressively over time, known as gradual drift. Identifying and mitigating drifts is one of the crucial activities during the monitoring phase of ML-enabled systems. This activity has attracted significant attention from the community, leading to various detection frameworks and approaches focusing on detecting and handling various types of drifts in ML-enabled systems \cite{zhengxin2023mlops, nelson2015evaluating, webb2016characterizing,bayram2022concept,vzliobaite2016overview, gama2014survey, gonccalves2014comparative,lu2018learning,mallick2022matchmaker,ackerman2021automatically,ackerman2020detection,poenaru2023maintaining,poenaru2022concept,lewis2022augur,yang2021cade,nahar2022collaboration}. Among the several drift detector frameworks proposed in the literature, a significant amount of work focuses on drift detection for classification tasks \cite{bayram2022concept}.

From a T\&E perspective, in the monitoring phase, there are several key considerations when it comes to detecting and mitigating drifts in ML-enabled systems:
\begin{itemize}
    \item Timely detection and mitigation of drift in ML-enabled systems are essential. The sooner a drift is identified and addressed, the less impact it will have on the performance of the ML-enabled system. Once the drift is detected, the practitioner must take the necessary steps to update the system and adapt it to the current operating environment.
    \item Drift detection frameworks are designed to quickly identify drifts in order to minimize the impact of drift and ensure the performance of machine learning (ML)-enabled systems in operational environments. However, this emphasis on swift detection may inadvertently compromise the framework's effectiveness, potentially leading to a higher number of false positives. This, in turn, can result in increased cost and effort in the post-deployment T\&E activities. 
    \item When a drift is identified, it is crucial to ensure that the corrective actions taken are thoroughly validated and do not unintentionally disrupt any existing functionalities. This validation process is essential to maintain the integrity and reliability of ML-enabled systems.
\end{itemize}

To identify and mitigate the impact of drift, practitioners employ drift detection frameworks. However, our survey of the literature suggests that various drift detection frameworks tend to suffer from a higher false-positive rate \cite{gama2014survey}. Additionally, the black-box nature of ML systems further exacerbates this. The opaque nature of model behavior, particularly DNN-based systems, makes it harder to understand and mitigate drifts \cite{gama2014survey,yang2021cade}.  

Given the nature of drifts, drift detection frameworks are mostly reactive in nature. From an operational standpoint, it would be ideal for detecting drift at the earliest possible stage, if not prior to its occurrence. In certain cases, the practitioner aims to leverage the operational environment's contextual information to equip themselves better to mitigate drifts that tend to occur periodically in a recurring fashion \cite{gama2014survey}.

Various approaches and tools have been developed to detect and mitigate drifts in ML-enabled systems. Although identifying drifts at the earliest is a good first step, understanding the severity of the risk posed by drifts can help practitioners determine appropriate corrective actions to mitigate the effects of drifts on model performance \cite{yang2021cade}. Furthermore, practitioners face challenges in adopting and utilizing monitoring tools in production environments \cite{nahar2022collaboration}, and in some cases, the practitioners do not actively pursue monitoring ML systems \cite{nahar2023meta}. The failure to undertake or perform monitoring activities by practitioners may result in performance degradation of ML models and eventually affect the performance of the ML-enabled software system.
 
\textbf{Reengineering in ML-enabled systems:}
Re-engineering, an important activity in the ML system lifecycle, is the process of systematically modifying an ML system to address any existing problems or to accommodate additional requirements. The primary objective of the re-engineering activity is to enhance the performance of ML systems. However, given the unique characteristics of ML systems, re-engineering is considered a challenging task. 

Jiang et al. investigated challenges in reengineering ML systems by analyzing 348 defects across 27 project repositories related to computer vision applications \cite{jiang2023challenges}. During re-engineering, practitioners may attempt to re-use an existing ML code and training dataset with the same platform configuration (as-is reuse), adapt an ML model for a different platform configuration, or introduce additional features to an existing implementation prior to its reuse. During this process, practitioners often encounter the challenge of being unable to execute the code and, subsequently, the execution, resulting in a crash. Another issue that is frequently experienced is that the program executes successfully but fails to meet the expected performance levels or the inferencing time does not meet the expectations.

Despite the availability of numerous open-source ML packages and models, the findings from this study suggest that caution must be exercised when forking projects from repositories. User forks are identified as a primary source of defects in reengineering ML systems. Additionally, practitioners encounter difficulties in executing ML projects from open-source repositories. While it is a common practice to provide instructions (README) for downloading, setting up, and executing the ML source code, in many instances, practitioners face difficulties in re-using an existing ML repository due to unclear or outdated README files. In some cases, repository owners fail to provide adequate instructions, leading practitioners to use incorrect configurations that subsequently affect the performance of ML-enabled systems. From a re-engineering standpoint, this issue is a major bottleneck in reusing ML-enabled systems. 

During the re-engineering process, defects may arise either from the data pipeline, training stage, modeling stage or from the environment. From their analysis, the authors observed that the training stage poses the most challenges in the re-engineering process, with most defects occurring at this stage. Defects introduced during the ML re-engineering process can broadly be classified into two categories: defects that trigger a crash and defects that don’t result in a crash but subtly impact the ML performance. From a T\&E perspective, the former is relatively easier to handle, as a system crash would require the teams to fix the issue before the ML system can be deployed. On the other hand, defects resulting in performance degradation pose a complex challenge and are harder to mitigate as they involve the tedious process of training and comparing over multiple iterations. This complexity is further exacerbated by the lack of well-established performance baselines to evaluate ML performance. The aforementioned challenges also extend to the debugging phase, requiring the development of new T\&E approaches to systematically detect and mitigate performance bugs during the model re-engineering process. 

Reusing ML components can significantly reduce the cost and effort required to build ML systems. Nevertheless, it presents its own set of challenges. Davis et al. investigated the challenges in reusing Deep Neural Network (DNN) models and categorized reusability into three distinct types: Conceptual Reuse, Adaptation Reuse, and Deployment Reuse \cite{davis2023reusing}. Nevertheless, practitioners face various engineering challenges when attempting to reuse existing ML-enabled systems. Model operationalization, performance debugging, and portability of DL operations are the three main DL re-engineering challenges \cite{jiang2023challenges}. The limitation in ML portability further exacerbates the reproducibility challenges in ML systems. As stated in \cite{li2022testing, islam2020repairing}, practitioners often face the challenge of training and operationalizing an ML model using a different ML framework than the one initially trained with or on a different hardware or a combination of both. Furthermore, most environment-related defects are triggered by API defects, and this is due to the limited support for backward compatibility provided by ML frameworks. In some cases, a subset of APIs may not be supported by certain frameworks while supported by the other frameworks. 

Another common approach among practitioners is the reuse of pre-trained ML models through adaptation. Using techniques such as transfer learning or knowledge distillation, pre-trained ML models are refined or adapted to perform tasks different from their original training. Furthermore, as discussed in the previous section, ML models may be transformed through quantization and pruning to operate in resource-constrained environments. Adapting ML models to operate on different platforms may adversely impact the model’s inference latency and prediction accuracy \cite{davis2023reusing}.

Overall, current re-engineering practices are predominantly ad-hoc and may prove inefficient, as they rely heavily on the domain knowledge and expertise of the practitioner. Engineering teams can benefit from standardizing the overall ML model re-engineering process to streamline the re-engineering process.

\textbf{Testing in the maintenance phase:}
An ML-enabled software system may contain one or more ML components. Based on their design, ML systems can be categorized into three groups -- single-model architecture, Pipeline architecture, and Hierarchical architecture \cite{li2022testing}. As the name suggests, the single-model architecture consists of one ML component responsible for performing intelligent tasks. On the other hand, pipeline architecture and Hierarchical architecture consist of more than one ML component. In pipeline architecture, tasks are delegated among the ML components, and each ML component is responsible for a particular subtask, whereas, in hierarchical architecture, ML components work together in tandem to accomplish intelligent tasks. 

The majority of existing work focuses on the test and evaluation of single-model architecture. Unlike the T\&E of single-model architecture, validating ML systems with a pipeline or hierarchical architecture is challenging. For example, consider an ML system with multiple ML components, and each ML component is responsible for a specific task. Assume that during the re-engineering process, one of the ML components is retrained to fix an issue identified earlier. This scenario necessitates a twofold T\&E criterion: First, the modified ML component must be validated individually, similar to the T\&E of single-model architecture. Second, T\&E efforts are needed to ensure the revised ML components do not adversely influence or impact the behavior of the rest of the ML components. In the current practice, practitioners may test the entire system (including all the ML components) despite the change being limited to one of the ML components \cite{li2022testing}. To the best of our knowledge, no work in the literature aims to address the challenges in T\&E of ML systems with either a pipeline or hierarchical architecture. Although the current practice aims to provide performance guarantees, it is inefficient and consumes a significant amount of time and effort. Furthermore, the current practice might be ad-hoc and thus limited in its applicability across different ML systems \cite{li2022testing}. Further research is warranted to address this gap and to develop approaches to perform effective T\&E for ML systems with multiple ML components in the deployment and re-engineering phases.

\textbf{Regression testing:} In traditional software systems, introducing additional functionalities or making changes or revisions to a software system typically involves changes to the source code. ML-enabled systems, on the other hand, encompass various artifacts, including training data, hyperparameters, ML algorithms, and metrics.  Therefore, revisions can take different forms in ML-enabled software systems, such as by adding or removing training data instances, modifying the ML model architecture, or changing the ML algorithm used in building the ML model. For example, a revision to an ML model can be accomplished by re-training the model with additional data or by re-tuning the model with a different set of hyperparameters, which adjusts the model’s internal weights. 

During the re-engineering phase, implementing new features or modifying existing features may involve modifying the source code in some cases, while in other cases, revisions may not involve changes to the source code. A revised model that achieves a higher score is considered to perform better. However, the revised model achieving a high accuracy does not guarantee the existing functionality is not adversely impacted. There is a possibility that the revised ML-enabled system may mispredict instances that were previously predicted correctly (before the latest release). Such inconsistencies in prediction between different versions of the ML system may impact the trustworthiness of the ML-enabled systems. A key challenge in the model re-engineering phase involves validating the changes in the revised versions of the ML software systems. 

From a T\&E perspective, it is essential to guarantee 
\begin{itemize}
    \item Newly introduced features or modified features perform as expected. 
    \item The revision does not disrupt any existing functionality.
\end{itemize}

Therefore, in addition to functional testing, regression testing is one of the widely followed testing approaches to validate changes in the software pipeline.  The main objective of regression testing is to ensure that the revisions to the system do not introduce new defects or adversely impact existing functionalities. However, performing regression testing for ML-enabled systems continues to be a challenge. Given the non-deterministic and data-intensive nature of ML systems, existing regression testing techniques do not suffice for ML systems. Existing studies from the literature suggest a lack of regression testing approaches that are well-suited for evaluating ML-enabled systems. Thus, new approaches and methodologies are needed to address this gap.

\textbf{Bug fixes in DNNs:} 
Islam et al. studied the challenges in repairing ML software systems, specifically the bug-fix patterns and the challenges faced by the practitioners in repairing DNNs \cite{islam2020repairing}. By analyzing the bug fixes from StackOverflow and GitHub, the authors classified the bug fix patterns into 15 types. Among these, fixing the dimension seems to be the most common bug-fix pattern in both GitHub and StackOverflow. For example, 23\% of the bugs (415) collected from StackOverflow were fixed by correcting the dimension of the input data or fixing layer dimensions. Furthermore, they identified that DNNs are brittle to changes in the input format. Other common fix patterns include Adding a layer to the DNN, modifying the loss function, and repairing the connections in the DNN.  

One of the challenges the practitioners encounter during the maintenance phase is to address or fix the failures triggered by incompatibility among different versions of ML libraries or frameworks. Lack of backward compatibility in ML frameworks can significantly impact maintenance activities. As \cite{islam2020repairing} reports, versioning-related bug fixes result in a higher maintenance cost. Notably, ML projects with prolonged durations are susceptible to versioning-related bugs due to frequently changing libraries. Consequently, failures or crashes may occur due to version or API mismatch between different framework versions, and it requires fixing API signatures to match the latest version.  Additionally, their study indicates more than 25\% of bug fixes introduce at least one new bug that differs from the original bug. Furthermore, they identified that changes to API signatures or API names are the most common root cause for the new bugs, and these bugs most likely trigger a Crash. Overall, versioning-related bugs could adversely impact the ML system maintenance activities, resulting in a cumbersome and costly affair \cite{jiang2023challenges, islam2020repairing}.

To recap, validating the performance of ML-enabled systems in the production environment is a challenging task compared to validating an ML model at the component level. Unlike traditional systems, maintenance activities in ML-enabled systems span across its lifecycle. Additionally, once deployed in a complex operational environment, the emergent behavior can result in unanticipated failures of ML-enabled systems. This necessitates developing proactive maintenance practices to guarantee the operational reliability and safety of ML-enabled systems and to prevent any undesirable behavior or harmful failures. To a larger extent, the current T\&E practices in the re-engineering remain ad-hoc, and standardizing processes could improve efficiency. Furthermore, new testing approaches are required to perform efficient testing, including regression testing, during the re-engineering phase.

\section{Conclusion} \label{conclusion}

ML-based software systems are increasingly adopted across domains. From a T\&E perspective, it is essential to guarantee that these systems behave as intended. This report presented best practices for T\&E across three stages of ML-enabled software systems: component stage, integration and deployment stage, and maintenance stage. From developing and testing the ML model as a component to seamlessly integrating with the rest of the components of the software system and finally to deploying and maintaining a reliable ML-enabled system continues to remain a challenge. The report discussed these challenges and current practices followed by practitioners at each stage. Furthermore, the report highlighted the need for new systematic approaches and metrics to address these challenges.

\section{Acknowledgments}

This material is based upon work supported, in whole or in part, by the U.S. Department of Defense through the Office of the Assistant Secretary of Defense for Research and Engineering (ASD(R\&E)) under Contract HQ003419D0003. The Systems Engineering Research Center (SERC) is a federally funded University Affiliated Research Center managed by Stevens Institute of Technology. Any views, opinions, findings and conclusions or recommendations expressed in this material are those of the author(s) and do not necessarily reflect the views of the United States Department of Defense nor ASD(R\&E).

\bibliographystyle{ACM-Reference-Format}
\bibliography{references}


\begin{thebibliography}{106}


\ifx \showCODEN    \undefined \def \showCODEN     #1{\unskip}     \fi
\ifx \showDOI      \undefined \def \showDOI       #1{#1}\fi
\ifx \showISBNx    \undefined \def \showISBNx     #1{\unskip}     \fi
\ifx \showISBNxiii \undefined \def \showISBNxiii  #1{\unskip}     \fi
\ifx \showISSN     \undefined \def \showISSN      #1{\unskip}     \fi
\ifx \showLCCN     \undefined \def \showLCCN      #1{\unskip}     \fi
\ifx \shownote     \undefined \def \shownote      #1{#1}          \fi
\ifx \showarticletitle \undefined \def \showarticletitle #1{#1}   \fi
\ifx \showURL      \undefined \def \showURL       {\relax}        \fi
\providecommand\bibfield[2]{#2}
\providecommand\bibinfo[2]{#2}
\providecommand\natexlab[1]{#1}
\providecommand\showeprint[2][]{arXiv:#2}

\bibitem[Cor({[n.\,d.]})]%
        {CoreML}
 \bibinfo{year}{[n.\,d.]}\natexlab{}.
\newblock \bibinfo{booktitle}{\emph{{Core ML} {A}pple {D}eveloper {D}ocumentation}}.
\newblock
\urldef\tempurl%
\url{https://developer.apple.com/documentation/coreml}
\showURL{%
\tempurl}


\bibitem[mll({[n.\,d.]})]%
        {mllib}
 \bibinfo{year}{[n.\,d.]}\natexlab{}.
\newblock \bibinfo{booktitle}{\emph{{ML}ib| {A}pache {S}park}}.
\newblock
\urldef\tempurl%
\url{https://spark.apache.org/mllib/}
\showURL{%
\tempurl}


\bibitem[mxn({[n.\,d.]})]%
        {mxnet}
 \bibinfo{year}{[n.\,d.]}\natexlab{}.
\newblock \bibinfo{booktitle}{\emph{mxnet-model-server}}.
\newblock
\urldef\tempurl%
\url{https://pypi.org/project/mxnet-model-server/}
\showURL{%
\tempurl}


\bibitem[Nvi({[n.\,d.]})]%
        {Nvidia}
 \bibinfo{year}{[n.\,d.]}\natexlab{}.
\newblock \bibinfo{booktitle}{\emph{{NVIDIA} {D}eep {L}earning {T}ensor{RT} {D}ocumentation}}.
\newblock
\urldef\tempurl%
\url{https://docs.nvidia.com/deeplearning/tensorrt/quick-start-guide/index.html}
\showURL{%
\tempurl}


\bibitem[ONN({[n.\,d.]})]%
        {ONNX}
 \bibinfo{year}{[n.\,d.]}\natexlab{}.
\newblock \bibinfo{booktitle}{\emph{{ONNX} | {H}ome}}.
\newblock
\urldef\tempurl%
\url{https://onnx.ai}
\showURL{%
\tempurl}


\bibitem[TFL({[n.\,d.]})]%
        {TFLite}
 \bibinfo{year}{[n.\,d.]}\natexlab{}.
\newblock \bibinfo{booktitle}{\emph{{Tensor{F}low {L}ite} {D}ocumentation}}.
\newblock
\urldef\tempurl%
\url{https://www.tensorflow.org/lite/guide}
\showURL{%
\tempurl}


\bibitem[caf({[n.\,d.]})]%
        {caffe2}
 \bibinfo{year}{[n.\,d.]}\natexlab{}.
\newblock \bibinfo{booktitle}{\emph{{W}hat is {C}affe2? | {C}affe2}}.
\newblock
\urldef\tempurl%
\url{https://caffe2.ai/docs/caffe-migration.html}
\showURL{%
\tempurl}


\bibitem[Ackerman et~al\mbox{.}(2020a)]%
        {ackerman2020detection}
\bibfield{author}{\bibinfo{person}{Samuel Ackerman}, \bibinfo{person}{Eitan Farchi}, \bibinfo{person}{Orna Raz}, \bibinfo{person}{Marcel Zalmanovici}, {and} \bibinfo{person}{Parijat Dube}.} \bibinfo{year}{2020}\natexlab{a}.
\newblock \showarticletitle{Detection of data drift and outliers affecting machine learning model performance over time}.
\newblock \bibinfo{journal}{\emph{arXiv preprint arXiv:2012.09258}} (\bibinfo{year}{2020}).
\newblock


\bibitem[Ackerman et~al\mbox{.}(2020b)]%
        {ackerman2020freaai}
\bibfield{author}{\bibinfo{person}{Samuel Ackerman}, \bibinfo{person}{Orna Raz}, {and} \bibinfo{person}{Marcel Zalmanovici}.} \bibinfo{year}{2020}\natexlab{b}.
\newblock \showarticletitle{FreaAI: Automated extraction of data slices to test machine learning models}. In \bibinfo{booktitle}{\emph{International Workshop on Engineering Dependable and Secure Machine Learning Systems}}. Springer, \bibinfo{pages}{67--83}.
\newblock


\bibitem[Ackerman et~al\mbox{.}(2021)]%
        {ackerman2021automatically}
\bibfield{author}{\bibinfo{person}{Samuel Ackerman}, \bibinfo{person}{Orna Raz}, \bibinfo{person}{Marcel Zalmanovici}, {and} \bibinfo{person}{Aviad Zlotnick}.} \bibinfo{year}{2021}\natexlab{}.
\newblock \showarticletitle{Automatically detecting data drift in machine learning classifiers}.
\newblock \bibinfo{journal}{\emph{arXiv preprint arXiv:2111.05672}} (\bibinfo{year}{2021}).
\newblock


\bibitem[Ahamed et~al\mbox{.}(2021)]%
        {ahamed2021attl}
\bibfield{author}{\bibinfo{person}{Sayyed~Farid Ahamed}, \bibinfo{person}{Priyanka Aggarwal}, \bibinfo{person}{Sachin Shetty}, \bibinfo{person}{Erin Lanus}, {and} \bibinfo{person}{Laura~J Freeman}.} \bibinfo{year}{2021}\natexlab{}.
\newblock \showarticletitle{ATTL: An Automated Targeted Transfer Learning with Deep Neural Networks}. In \bibinfo{booktitle}{\emph{2021 IEEE Global Communications Conference (GLOBECOM)}}. IEEE, \bibinfo{pages}{1--7}.
\newblock


\bibitem[Ahuja et~al\mbox{.}(2022)]%
        {ahuja2022testing}
\bibfield{author}{\bibinfo{person}{Mohit~Kumar Ahuja}, \bibinfo{person}{Arnaud Gotlieb}, {and} \bibinfo{person}{Helge Spieker}.} \bibinfo{year}{2022}\natexlab{}.
\newblock \showarticletitle{Testing Deep Learning Models: A First Comparative Study of Multiple Testing Techniques}. In \bibinfo{booktitle}{\emph{2022 IEEE International Conference on Software Testing, Verification and Validation Workshops (ICSTW)}}. IEEE, \bibinfo{pages}{130--137}.
\newblock


\bibitem[Al-Azani and Hassine(2017)]%
        {al2017validation}
\bibfield{author}{\bibinfo{person}{Sadam Al-Azani} {and} \bibinfo{person}{Jameleddine Hassine}.} \bibinfo{year}{2017}\natexlab{}.
\newblock \showarticletitle{Validation of machine learning classifiers using metamorphic testing and feature selection techniques}. In \bibinfo{booktitle}{\emph{Multi-disciplinary Trends in Artificial Intelligence: 11th International Workshop, MIWAI 2017, Gadong, Brunei, November 20-22, 2017, Proceedings 11}}. Springer, \bibinfo{pages}{77--91}.
\newblock


\bibitem[Asyrofi et~al\mbox{.}(2020)]%
        {asyrofi2020crossasr}
\bibfield{author}{\bibinfo{person}{Muhammad~Hilmi Asyrofi}, \bibinfo{person}{Ferdian Thung}, \bibinfo{person}{David Lo}, {and} \bibinfo{person}{Lingxiao Jiang}.} \bibinfo{year}{2020}\natexlab{}.
\newblock \showarticletitle{Crossasr: Efficient differential testing of automatic speech recognition via text-to-speech}. In \bibinfo{booktitle}{\emph{2020 IEEE International Conference on Software Maintenance and Evolution (ICSME)}}. IEEE, \bibinfo{pages}{640--650}.
\newblock


\bibitem[Baier et~al\mbox{.}(2019)]%
        {baier2019challenges}
\bibfield{author}{\bibinfo{person}{Lucas Baier}, \bibinfo{person}{Fabian J{\"o}hren}, {and} \bibinfo{person}{Stefan Seebacher}.} \bibinfo{year}{2019}\natexlab{}.
\newblock \showarticletitle{Challenges in the Deployment and Operation of Machine Learning in Practice.}. In \bibinfo{booktitle}{\emph{ECIS}}, Vol.~\bibinfo{volume}{1}.
\newblock


\bibitem[Barash et~al\mbox{.}(2019)]%
        {barash2019bridging}
\bibfield{author}{\bibinfo{person}{Guy Barash}, \bibinfo{person}{Eitan Farchi}, \bibinfo{person}{Ilan Jayaraman}, \bibinfo{person}{Orna Raz}, \bibinfo{person}{Rachel Tzoref-Brill}, {and} \bibinfo{person}{Marcel Zalmanovici}.} \bibinfo{year}{2019}\natexlab{}.
\newblock \showarticletitle{Bridging the gap between ml solutions and their business requirements using feature interactions}. In \bibinfo{booktitle}{\emph{Proceedings of the 2019 27th ACM Joint Meeting on European Software Engineering Conference and Symposium on the Foundations of Software Engineering}}. \bibinfo{pages}{1048--1058}.
\newblock


\bibitem[Bayram et~al\mbox{.}(2022)]%
        {bayram2022concept}
\bibfield{author}{\bibinfo{person}{Firas Bayram}, \bibinfo{person}{Bestoun~S Ahmed}, {and} \bibinfo{person}{Andreas Kassler}.} \bibinfo{year}{2022}\natexlab{}.
\newblock \showarticletitle{From concept drift to model degradation: An overview on performance-aware drift detectors}.
\newblock \bibinfo{journal}{\emph{Knowledge-Based Systems}}  \bibinfo{volume}{245} (\bibinfo{year}{2022}), \bibinfo{pages}{108632}.
\newblock


\bibitem[Buitinck et~al\mbox{.}(2013)]%
        {sklearn_api}
\bibfield{author}{\bibinfo{person}{Lars Buitinck}, \bibinfo{person}{Gilles Louppe}, \bibinfo{person}{Mathieu Blondel}, \bibinfo{person}{Fabian Pedregosa}, \bibinfo{person}{Andreas Mueller}, \bibinfo{person}{Olivier Grisel}, \bibinfo{person}{Vlad Niculae}, \bibinfo{person}{Peter Prettenhofer}, \bibinfo{person}{Alexandre Gramfort}, \bibinfo{person}{Jaques Grobler}, \bibinfo{person}{Robert Layton}, \bibinfo{person}{Jake VanderPlas}, \bibinfo{person}{Arnaud Joly}, \bibinfo{person}{Brian Holt}, {and} \bibinfo{person}{Ga{\"{e}}l Varoquaux}.} \bibinfo{year}{2013}\natexlab{}.
\newblock \showarticletitle{{API} design for machine learning software: experiences from the scikit-learn project}. In \bibinfo{booktitle}{\emph{ECML PKDD Workshop: Languages for Data Mining and Machine Learning}}. \bibinfo{pages}{108--122}.
\newblock


\bibitem[Chandrasekaran(2021)]%
        {chandrasekaran2021testing}
\bibfield{author}{\bibinfo{person}{Jaganmohan Chandrasekaran}.} \bibinfo{year}{2021}\natexlab{}.
\newblock \bibinfo{booktitle}{\emph{Testing Artificial Intelligence-Based Software Systems}}.
\newblock \bibinfo{publisher}{The University of Texas at Arlington}.
\newblock


\bibitem[Chandrasekaran et~al\mbox{.}(2017)]%
        {chandrasekaran2017applying}
\bibfield{author}{\bibinfo{person}{Jaganmohan Chandrasekaran}, \bibinfo{person}{Huadong Feng}, \bibinfo{person}{Yu Lei}, \bibinfo{person}{D~Richard Kuhn}, {and} \bibinfo{person}{Raghu Kacker}.} \bibinfo{year}{2017}\natexlab{}.
\newblock \showarticletitle{Applying combinatorial testing to data mining algorithms}. In \bibinfo{booktitle}{\emph{2017 IEEE International Conference on Software Testing, Verification and Validation Workshops (ICSTW)}}. IEEE, \bibinfo{pages}{253--261}.
\newblock


\bibitem[Chandrasekaran et~al\mbox{.}({[n.\,d.]})]%
        {CCMagazine}
\bibfield{author}{\bibinfo{person}{Jaganmohan Chandrasekaran}, \bibinfo{person}{Erin Lanus}, \bibinfo{person}{Tyler Cody}, \bibinfo{person}{Laura Freeman}, \bibinfo{person}{Raghu Kacker}, \bibinfo{person}{M~S Raunak}, {and} \bibinfo{person}{D.~Richard Kuhn}.} \bibinfo{year}{[n.\,d.]}\natexlab{}.
\newblock \bibinfo{title}{Leveraging {C}ombinatorial {C}overage in {ML} {P}roduct {L}ifecycle}.  (\bibinfo{year}{[n.\,d.]}).
\newblock
\newblock
\shownote{{U}nder {R}eview}.


\bibitem[Chandrasekaran et~al\mbox{.}(2021a)]%
        {chandrasekaran2021combinatorial}
\bibfield{author}{\bibinfo{person}{Jaganmohan Chandrasekaran}, \bibinfo{person}{Yu Lei}, \bibinfo{person}{Raghu Kacker}, {and} \bibinfo{person}{D~Richard Kuhn}.} \bibinfo{year}{2021}\natexlab{a}.
\newblock \showarticletitle{A combinatorial approach to explaining image classifiers}. In \bibinfo{booktitle}{\emph{2021 IEEE International Conference on Software Testing, Verification and Validation Workshops (ICSTW)}}. IEEE, \bibinfo{pages}{35--43}.
\newblock


\bibitem[Chandrasekaran et~al\mbox{.}(2021b)]%
        {chandrasekaran2021combinatorialADS}
\bibfield{author}{\bibinfo{person}{Jaganmohan Chandrasekaran}, \bibinfo{person}{Yu Lei}, \bibinfo{person}{Raghu Kacker}, {and} \bibinfo{person}{D~Richard Kuhn}.} \bibinfo{year}{2021}\natexlab{b}.
\newblock \showarticletitle{A combinatorial approach to testing deep neural network-based autonomous driving systems}. In \bibinfo{booktitle}{\emph{2021 IEEE International Conference on Software Testing, Verification and Validation Workshops (ICSTW)}}. IEEE, \bibinfo{pages}{57--66}.
\newblock


\bibitem[Chen and Das(2023)]%
        {chen2023ai}
\bibfield{author}{\bibinfo{person}{Pin-Yu Chen} {and} \bibinfo{person}{Payel Das}.} \bibinfo{year}{2023}\natexlab{}.
\newblock \showarticletitle{{A}{I} Maintenance: A {R}obustness {P}erspective}.
\newblock \bibinfo{journal}{\emph{Computer}} \bibinfo{volume}{56}, \bibinfo{number}{2} (\bibinfo{year}{2023}), \bibinfo{pages}{48--56}.
\newblock


\bibitem[Chen et~al\mbox{.}(2019)]%
        {chen2019variable}
\bibfield{author}{\bibinfo{person}{Yanshan Chen}, \bibinfo{person}{Ziyuan Wang}, \bibinfo{person}{Dong Wang}, \bibinfo{person}{Chunrong Fang}, {and} \bibinfo{person}{Zhenyu Chen}.} \bibinfo{year}{2019}\natexlab{}.
\newblock \showarticletitle{Variable strength combinatorial testing for deep neural networks}. In \bibinfo{booktitle}{\emph{2019 IEEE International Conference on Software Testing, Verification and Validation Workshops (ICSTW)}}. IEEE, \bibinfo{pages}{281--284}.
\newblock


\bibitem[Chen et~al\mbox{.}(2020)]%
        {chen2020comprehensive}
\bibfield{author}{\bibinfo{person}{Zhenpeng Chen}, \bibinfo{person}{Yanbin Cao}, \bibinfo{person}{Yuanqiang Liu}, \bibinfo{person}{Haoyu Wang}, \bibinfo{person}{Tao Xie}, {and} \bibinfo{person}{Xuanzhe Liu}.} \bibinfo{year}{2020}\natexlab{}.
\newblock \showarticletitle{A comprehensive study on challenges in deploying deep learning based software}. In \bibinfo{booktitle}{\emph{Proceedings of the 28th ACM Joint Meeting on European Software Engineering Conference and Symposium on the Foundations of Software Engineering}}. \bibinfo{pages}{750--762}.
\newblock


\bibitem[Chen et~al\mbox{.}(2021)]%
        {chen2021empirical}
\bibfield{author}{\bibinfo{person}{Zhenpeng Chen}, \bibinfo{person}{Huihan Yao}, \bibinfo{person}{Yiling Lou}, \bibinfo{person}{Yanbin Cao}, \bibinfo{person}{Yuanqiang Liu}, \bibinfo{person}{Haoyu Wang}, {and} \bibinfo{person}{Xuanzhe Liu}.} \bibinfo{year}{2021}\natexlab{}.
\newblock \showarticletitle{An {E}mpirical {S}tudy on {D}eployment {F}aults of {D}eep {L}earning {B]ased {M}obile {A}pplications}}. In \bibinfo{booktitle}{\emph{2021 IEEE/ACM 43rd International Conference on Software Engineering (ICSE)}}. IEEE, \bibinfo{pages}{674--685}.
\newblock


\bibitem[Cody et~al\mbox{.}(2022)]%
        {cody2022systematic}
\bibfield{author}{\bibinfo{person}{Tyler Cody}, \bibinfo{person}{Erin Lanus}, \bibinfo{person}{Daniel~D Doyle}, {and} \bibinfo{person}{Laura Freeman}.} \bibinfo{year}{2022}\natexlab{}.
\newblock \showarticletitle{Systematic training and testing for machine learning using combinatorial interaction testing}. In \bibinfo{booktitle}{\emph{2022 IEEE International Conference on Software Testing, Verification and Validation Workshops (ICSTW)}}. IEEE, \bibinfo{pages}{102--109}.
\newblock


\bibitem[Croce et~al\mbox{.}(2020)]%
        {croce2020robustbench}
\bibfield{author}{\bibinfo{person}{Francesco Croce}, \bibinfo{person}{Maksym Andriushchenko}, \bibinfo{person}{Vikash Sehwag}, \bibinfo{person}{Edoardo Debenedetti}, \bibinfo{person}{Nicolas Flammarion}, \bibinfo{person}{Mung Chiang}, \bibinfo{person}{Prateek Mittal}, {and} \bibinfo{person}{Matthias Hein}.} \bibinfo{year}{2020}\natexlab{}.
\newblock \showarticletitle{Robustbench: a standardized adversarial robustness benchmark}.
\newblock \bibinfo{journal}{\emph{arXiv preprint arXiv:2010.09670}} (\bibinfo{year}{2020}).
\newblock


\bibitem[Davis et~al\mbox{.}(2023)]%
        {davis2023reusing}
\bibfield{author}{\bibinfo{person}{James~C Davis}, \bibinfo{person}{Purvish Jajal}, \bibinfo{person}{Wenxin Jiang}, \bibinfo{person}{Taylor~R Schorlemmer}, \bibinfo{person}{Nicholas Synovic}, {and} \bibinfo{person}{George~K Thiruvathukal}.} \bibinfo{year}{2023}\natexlab{}.
\newblock \showarticletitle{Reusing Deep Learning Models: Challenges and Directions in Software Engineering}.
\newblock  (\bibinfo{year}{2023}).
\newblock


\bibitem[Deng et~al\mbox{.}(2022)]%
        {deng2022fuzzing}
\bibfield{author}{\bibinfo{person}{Yinlin Deng}, \bibinfo{person}{Chenyuan Yang}, \bibinfo{person}{Anjiang Wei}, {and} \bibinfo{person}{Lingming Zhang}.} \bibinfo{year}{2022}\natexlab{}.
\newblock \showarticletitle{Fuzzing {D}eep-learning libraries via {A}utomated {R}elational {API} {I}nference}. In \bibinfo{booktitle}{\emph{Proceedings of the 30th ACM Joint European Software Engineering Conference and Symposium on the Foundations of Software Engineering}}. \bibinfo{pages}{44--56}.
\newblock


\bibitem[Ding et~al\mbox{.}(2017)]%
        {ding2017validating}
\bibfield{author}{\bibinfo{person}{Junhua Ding}, \bibinfo{person}{Xiaojun Kang}, {and} \bibinfo{person}{Xin-Hua Hu}.} \bibinfo{year}{2017}\natexlab{}.
\newblock \showarticletitle{Validating a deep learning framework by metamorphic testing}. In \bibinfo{booktitle}{\emph{2017 IEEE/ACM 2nd International Workshop on Metamorphic Testing (MET)}}. IEEE, \bibinfo{pages}{28--34}.
\newblock


\bibitem[Du et~al\mbox{.}(2018)]%
        {du2018deepcruiser}
\bibfield{author}{\bibinfo{person}{Xiaoning Du}, \bibinfo{person}{Xiaofei Xie}, \bibinfo{person}{Yi Li}, \bibinfo{person}{Lei Ma}, \bibinfo{person}{Jianjun Zhao}, {and} \bibinfo{person}{Yang Liu}.} \bibinfo{year}{2018}\natexlab{}.
\newblock \showarticletitle{Deepcruiser: Automated guided testing for stateful deep learning systems}.
\newblock \bibinfo{journal}{\emph{arXiv preprint arXiv:1812.05339}} (\bibinfo{year}{2018}).
\newblock


\bibitem[Dwarakanath et~al\mbox{.}(2018)]%
        {dwarakanath2018identifying}
\bibfield{author}{\bibinfo{person}{Anurag Dwarakanath}, \bibinfo{person}{Manish Ahuja}, \bibinfo{person}{Samarth Sikand}, \bibinfo{person}{Raghotham~M Rao}, \bibinfo{person}{RP~Jagadeesh~Chandra Bose}, \bibinfo{person}{Neville Dubash}, {and} \bibinfo{person}{Sanjay Podder}.} \bibinfo{year}{2018}\natexlab{}.
\newblock \showarticletitle{Identifying implementation bugs in machine learning based image classifiers using metamorphic testing}. In \bibinfo{booktitle}{\emph{Proceedings of the 27th ACM SIGSOFT international symposium on software testing and analysis}}. \bibinfo{pages}{118--128}.
\newblock


\bibitem[Dwork(2008)]%
        {dwork2008differential}
\bibfield{author}{\bibinfo{person}{Cynthia Dwork}.} \bibinfo{year}{2008}\natexlab{}.
\newblock \showarticletitle{Differential privacy: A survey of results}. In \bibinfo{booktitle}{\emph{International conference on theory and applications of models of computation}}. Springer, \bibinfo{pages}{1--19}.
\newblock


\bibitem[Dwork et~al\mbox{.}(2006)]%
        {dwork2006our}
\bibfield{author}{\bibinfo{person}{Cynthia Dwork}, \bibinfo{person}{Krishnaram Kenthapadi}, \bibinfo{person}{Frank McSherry}, \bibinfo{person}{Ilya Mironov}, {and} \bibinfo{person}{Moni Naor}.} \bibinfo{year}{2006}\natexlab{}.
\newblock \showarticletitle{Our data, ourselves: Privacy via distributed noise generation}. In \bibinfo{booktitle}{\emph{Advances in Cryptology-EUROCRYPT 2006: 24th Annual International Conference on the Theory and Applications of Cryptographic Techniques, St. Petersburg, Russia, May 28-June 1, 2006. Proceedings 25}}. Springer, \bibinfo{pages}{486--503}.
\newblock


\bibitem[Eibe et~al\mbox{.}(2016)]%
        {eibe2016weka}
\bibfield{author}{\bibinfo{person}{Frank Eibe}, \bibinfo{person}{Mark~A Hall}, {and} \bibinfo{person}{Ian~H Witten}.} \bibinfo{year}{2016}\natexlab{}.
\newblock \showarticletitle{The WEKA workbench. Online appendix for data mining: practical machine learning tools and techniques}.
\newblock In \bibinfo{booktitle}{\emph{Morgan Kaufmann}}. \bibinfo{publisher}{Morgan Kaufmann Publishers San Francisco, California}.
\newblock


\bibitem[Gama et~al\mbox{.}(2014)]%
        {gama2014survey}
\bibfield{author}{\bibinfo{person}{Jo{\~a}o Gama}, \bibinfo{person}{Indr{\.e} {\v{Z}}liobait{\.e}}, \bibinfo{person}{Albert Bifet}, \bibinfo{person}{Mykola Pechenizkiy}, {and} \bibinfo{person}{Abdelhamid Bouchachia}.} \bibinfo{year}{2014}\natexlab{}.
\newblock \showarticletitle{A survey on concept drift adaptation}.
\newblock \bibinfo{journal}{\emph{ACM computing surveys (CSUR)}} \bibinfo{volume}{46}, \bibinfo{number}{4} (\bibinfo{year}{2014}), \bibinfo{pages}{1--37}.
\newblock


\bibitem[Giray(2021)]%
        {giray2021software}
\bibfield{author}{\bibinfo{person}{G{\"o}rkem Giray}.} \bibinfo{year}{2021}\natexlab{}.
\newblock \showarticletitle{A software engineering perspective on engineering machine learning systems: State of the art and challenges}.
\newblock \bibinfo{journal}{\emph{Journal of Systems and Software}}  \bibinfo{volume}{180} (\bibinfo{year}{2021}), \bibinfo{pages}{111031}.
\newblock


\bibitem[Gladisch et~al\mbox{.}(2020)]%
        {gladisch2020leveraging}
\bibfield{author}{\bibinfo{person}{Christoph Gladisch}, \bibinfo{person}{Christian Heinzemann}, \bibinfo{person}{Martin Herrmann}, {and} \bibinfo{person}{Matthias Woehrle}.} \bibinfo{year}{2020}\natexlab{}.
\newblock \showarticletitle{Leveraging combinatorial testing for safety-critical computer vision datasets}. In \bibinfo{booktitle}{\emph{Proceedings of the IEEE/CVF Conference on Computer Vision and Pattern Recognition Workshops}}. \bibinfo{pages}{324--325}.
\newblock


\bibitem[Gon{\c{c}}alves~Jr et~al\mbox{.}(2014)]%
        {gonccalves2014comparative}
\bibfield{author}{\bibinfo{person}{Paulo~M Gon{\c{c}}alves~Jr}, \bibinfo{person}{Silas~GT de Carvalho~Santos}, \bibinfo{person}{Roberto~SM Barros}, {and} \bibinfo{person}{Davi~CL Vieira}.} \bibinfo{year}{2014}\natexlab{}.
\newblock \showarticletitle{A comparative study on concept drift detectors}.
\newblock \bibinfo{journal}{\emph{Expert Systems with Applications}} \bibinfo{volume}{41}, \bibinfo{number}{18} (\bibinfo{year}{2014}), \bibinfo{pages}{8144--8156}.
\newblock


\bibitem[Granja-Alvarez and Barranco-Garc{\'\i}a(1997)]%
        {granja1997method}
\bibfield{author}{\bibinfo{person}{Juan~Carlos Granja-Alvarez} {and} \bibinfo{person}{Manuel~Jos{\'e} Barranco-Garc{\'\i}a}.} \bibinfo{year}{1997}\natexlab{}.
\newblock \showarticletitle{A method for estimating maintenance cost in a software project: a case study}.
\newblock \bibinfo{journal}{\emph{Journal of Software Maintenance: Research and Practice}} \bibinfo{volume}{9}, \bibinfo{number}{3} (\bibinfo{year}{1997}), \bibinfo{pages}{161--175}.
\newblock


\bibitem[Guo et~al\mbox{.}(2018)]%
        {guo2018dlfuzz}
\bibfield{author}{\bibinfo{person}{Jianmin Guo}, \bibinfo{person}{Yu Jiang}, \bibinfo{person}{Yue Zhao}, \bibinfo{person}{Quan Chen}, {and} \bibinfo{person}{Jiaguang Sun}.} \bibinfo{year}{2018}\natexlab{}.
\newblock \showarticletitle{Dlfuzz: Differential fuzzing testing of deep learning systems}. In \bibinfo{booktitle}{\emph{Proceedings of the 2018 26th ACM Joint Meeting on European Software Engineering Conference and Symposium on the Foundations of Software Engineering}}. \bibinfo{pages}{739--743}.
\newblock


\bibitem[Guo et~al\mbox{.}(2019)]%
        {guo2019empirical}
\bibfield{author}{\bibinfo{person}{Qianyu Guo}, \bibinfo{person}{Sen Chen}, \bibinfo{person}{Xiaofei Xie}, \bibinfo{person}{Lei Ma}, \bibinfo{person}{Qiang Hu}, \bibinfo{person}{Hongtao Liu}, \bibinfo{person}{Yang Liu}, \bibinfo{person}{Jianjun Zhao}, {and} \bibinfo{person}{Xiaohong Li}.} \bibinfo{year}{2019}\natexlab{}.
\newblock \showarticletitle{An empirical study towards characterizing deep learning development and deployment across different frameworks and platforms}. In \bibinfo{booktitle}{\emph{2019 34th IEEE/ACM International Conference on Automated Software Engineering (ASE)}}. IEEE, \bibinfo{pages}{810--822}.
\newblock


\bibitem[Hall et~al\mbox{.}(2009)]%
        {hall2009weka}
\bibfield{author}{\bibinfo{person}{Mark Hall}, \bibinfo{person}{Eibe Frank}, \bibinfo{person}{Geoffrey Holmes}, \bibinfo{person}{Bernhard Pfahringer}, \bibinfo{person}{Peter Reutemann}, {and} \bibinfo{person}{Ian~H Witten}.} \bibinfo{year}{2009}\natexlab{}.
\newblock \showarticletitle{The WEKA data mining software: an update}.
\newblock \bibinfo{journal}{\emph{ACM SIGKDD explorations newsletter}} \bibinfo{volume}{11}, \bibinfo{number}{1} (\bibinfo{year}{2009}), \bibinfo{pages}{10--18}.
\newblock


\bibitem[Harel-Canada et~al\mbox{.}(2020)]%
        {harel2020neuron}
\bibfield{author}{\bibinfo{person}{Fabrice Harel-Canada}, \bibinfo{person}{Lingxiao Wang}, \bibinfo{person}{Muhammad~Ali Gulzar}, \bibinfo{person}{Quanquan Gu}, {and} \bibinfo{person}{Miryung Kim}.} \bibinfo{year}{2020}\natexlab{}.
\newblock \showarticletitle{Is neuron coverage a meaningful measure for testing deep neural networks?}. In \bibinfo{booktitle}{\emph{Proceedings of the 28th ACM Joint Meeting on European Software Engineering Conference and Symposium on the Foundations of Software Engineering}}. \bibinfo{pages}{851--862}.
\newblock


\bibitem[Herbold and Haar(2022)]%
        {herbold2022smoke}
\bibfield{author}{\bibinfo{person}{Steffen Herbold} {and} \bibinfo{person}{Tobias Haar}.} \bibinfo{year}{2022}\natexlab{}.
\newblock \showarticletitle{Smoke testing for machine learning: simple tests to discover severe bugs}.
\newblock \bibinfo{journal}{\emph{Empirical Software Engineering}} \bibinfo{volume}{27}, \bibinfo{number}{2} (\bibinfo{year}{2022}), \bibinfo{pages}{45}.
\newblock


\bibitem[Herbold and Tunkel(2023)]%
        {herbold2023differential}
\bibfield{author}{\bibinfo{person}{Steffen Herbold} {and} \bibinfo{person}{Steffen Tunkel}.} \bibinfo{year}{2023}\natexlab{}.
\newblock \showarticletitle{Differential testing for machine learning: an analysis for classification algorithms beyond deep learning}.
\newblock \bibinfo{journal}{\emph{Empirical Software Engineering}} \bibinfo{volume}{28}, \bibinfo{number}{2} (\bibinfo{year}{2023}), \bibinfo{pages}{34}.
\newblock


\bibitem[Hu et~al\mbox{.}(2023)]%
        {hu2023towards}
\bibfield{author}{\bibinfo{person}{Qiang Hu}, \bibinfo{person}{Yuejun Guo}, \bibinfo{person}{Maxime Cordy}, \bibinfo{person}{Xiaofei Xie}, \bibinfo{person}{Wei Ma}, \bibinfo{person}{Mike Papadakis}, {and} \bibinfo{person}{Yves Le~Traon}.} \bibinfo{year}{2023}\natexlab{}.
\newblock \showarticletitle{Towards Understanding Model Quantization for Reliable Deep Neural Network Deployment}. In \bibinfo{booktitle}{\emph{2023 IEEE/ACM 2nd International Conference on AI Engineering--Software Engineering for AI (CAIN)}}. IEEE, \bibinfo{pages}{56--67}.
\newblock


\bibitem[Hutchinson et~al\mbox{.}(2022)]%
        {hutchinson2022evaluation}
\bibfield{author}{\bibinfo{person}{Ben Hutchinson}, \bibinfo{person}{Negar Rostamzadeh}, \bibinfo{person}{Christina Greer}, \bibinfo{person}{Katherine Heller}, {and} \bibinfo{person}{Vinodkumar Prabhakaran}.} \bibinfo{year}{2022}\natexlab{}.
\newblock \showarticletitle{Evaluation gaps in machine learning practice}. In \bibinfo{booktitle}{\emph{Proceedings of the 2022 ACM Conference on Fairness, Accountability, and Transparency}}. \bibinfo{pages}{1859--1876}.
\newblock


\bibitem[Islam et~al\mbox{.}(2020)]%
        {islam2020repairing}
\bibfield{author}{\bibinfo{person}{Md~Johirul Islam}, \bibinfo{person}{Rangeet Pan}, \bibinfo{person}{Giang Nguyen}, {and} \bibinfo{person}{Hridesh Rajan}.} \bibinfo{year}{2020}\natexlab{}.
\newblock \showarticletitle{Repairing deep neural networks: Fix patterns and challenges}. In \bibinfo{booktitle}{\emph{Proceedings of the ACM/IEEE 42nd International Conference on Software Engineering}}. \bibinfo{pages}{1135--1146}.
\newblock


\bibitem[Jajal et~al\mbox{.}(2023)]%
        {jajal2023analysis}
\bibfield{author}{\bibinfo{person}{Purvish Jajal}, \bibinfo{person}{Wenxin Jiang}, \bibinfo{person}{Arav Tewari}, \bibinfo{person}{Joseph Woo}, \bibinfo{person}{Yung-Hsiang Lu}, \bibinfo{person}{George~K Thiruvathukal}, {and} \bibinfo{person}{James~C Davis}.} \bibinfo{year}{2023}\natexlab{}.
\newblock \showarticletitle{Analysis of Failures and Risks in Deep Learning Model Converters: A Case Study in the ONNX Ecosystem}.
\newblock \bibinfo{journal}{\emph{arXiv preprint arXiv:2303.17708}} (\bibinfo{year}{2023}).
\newblock


\bibitem[Jiang et~al\mbox{.}(2023)]%
        {jiang2023challenges}
\bibfield{author}{\bibinfo{person}{Wenxin Jiang}, \bibinfo{person}{Vishnu Banna}, \bibinfo{person}{Naveen Vivek}, \bibinfo{person}{Abhinav Goel}, \bibinfo{person}{Nicholas Synovic}, \bibinfo{person}{George~K Thiruvathukal}, {and} \bibinfo{person}{James~C Davis}.} \bibinfo{year}{2023}\natexlab{}.
\newblock \showarticletitle{Challenges and {P}ractices of {D}eep {L}earning {M}odel {R}eengineering: A {C}ase {S}tudy on {C}omputer {V}ision}.
\newblock \bibinfo{journal}{\emph{arXiv preprint arXiv:2303.07476}} (\bibinfo{year}{2023}).
\newblock


\bibitem[Katragadda et~al\mbox{.}(2022)]%
        {katragadda2022active}
\bibfield{author}{\bibinfo{person}{Sai~Prathyush Katragadda}, \bibinfo{person}{Tyler Cody}, \bibinfo{person}{Peter Beling}, {and} \bibinfo{person}{Laura Freeman}.} \bibinfo{year}{2022}\natexlab{}.
\newblock \showarticletitle{Active learning with combinatorial coverage}. In \bibinfo{booktitle}{\emph{2022 21st IEEE International Conference on Machine Learning and Applications (ICMLA)}}. IEEE, \bibinfo{pages}{1129--1136}.
\newblock


\bibitem[Khadka et~al\mbox{.}(2023)]%
        {khadka2023synthetic}
\bibfield{author}{\bibinfo{person}{Krishna Khadka}, \bibinfo{person}{Jaganmohan Chandrasekaran}, \bibinfo{person}{Yu Lei}, \bibinfo{person}{Raghu~N Kacker}, {and} \bibinfo{person}{D~Richard Kuhn}.} \bibinfo{year}{2023}\natexlab{}.
\newblock \showarticletitle{Synthetic Data Generation Using Combinatorial Testing and Variational Autoencoder}. In \bibinfo{booktitle}{\emph{2023 IEEE International Conference on Software Testing, Verification and Validation Workshops (ICSTW)}}. IEEE, \bibinfo{pages}{228--236}.
\newblock


\bibitem[Kim et~al\mbox{.}(2019)]%
        {kim2019guiding}
\bibfield{author}{\bibinfo{person}{Jinhan Kim}, \bibinfo{person}{Robert Feldt}, {and} \bibinfo{person}{Shin Yoo}.} \bibinfo{year}{2019}\natexlab{}.
\newblock \showarticletitle{Guiding deep learning system testing using surprise adequacy}. In \bibinfo{booktitle}{\emph{2019 IEEE/ACM 41st International Conference on Software Engineering (ICSE)}}. IEEE, \bibinfo{pages}{1039--1049}.
\newblock


\bibitem[{Kuhn} and {Max}(2008)]%
        {caret}
\bibfield{author}{\bibinfo{person}{{Kuhn}} {and} \bibinfo{person}{{Max}}.} \bibinfo{year}{2008}\natexlab{}.
\newblock \showarticletitle{Building Predictive Models in R Using the caret Package}.
\newblock \bibinfo{journal}{\emph{Journal of Statistical Software}} \bibinfo{volume}{28}, \bibinfo{number}{5} (\bibinfo{year}{2008}), \bibinfo{pages}{1–26}.
\newblock
\urldef\tempurl%
\url{https://doi.org/10.18637/jss.v028.i05}
\showDOI{\tempurl}


\bibitem[Kuhn et~al\mbox{.}(2010)]%
        {kuhn2010practical}
\bibfield{author}{\bibinfo{person}{D~Richard Kuhn}, \bibinfo{person}{Raghu~N Kacker}, \bibinfo{person}{Yu Lei}, {et~al\mbox{.}}} \bibinfo{year}{2010}\natexlab{}.
\newblock \showarticletitle{Practical combinatorial testing}.
\newblock \bibinfo{journal}{\emph{NIST special Publication}} \bibinfo{volume}{800}, \bibinfo{number}{142} (\bibinfo{year}{2010}), \bibinfo{pages}{142}.
\newblock


\bibitem[Kuhn et~al\mbox{.}(2020)]%
        {kuhn2020combinatorial}
\bibfield{author}{\bibinfo{person}{D~Richard Kuhn}, \bibinfo{person}{Raghu~N Kacker}, \bibinfo{person}{Yu Lei}, {and} \bibinfo{person}{Dimitris~E Simos}.} \bibinfo{year}{2020}\natexlab{}.
\newblock \showarticletitle{Combinatorial methods for explainable AI}. In \bibinfo{booktitle}{\emph{2020 IEEE International Conference on Software Testing, Verification and Validation Workshops (ICSTW)}}. IEEE, \bibinfo{pages}{167--170}.
\newblock


\bibitem[Kuhn et~al\mbox{.}(2004)]%
        {kuhn2004software}
\bibfield{author}{\bibinfo{person}{D~Richard Kuhn}, \bibinfo{person}{Dolores~R Wallace}, {and} \bibinfo{person}{Albert~M Gallo}.} \bibinfo{year}{2004}\natexlab{}.
\newblock \showarticletitle{Software fault interactions and implications for software testing}.
\newblock \bibinfo{journal}{\emph{IEEE transactions on software engineering}} \bibinfo{volume}{30}, \bibinfo{number}{6} (\bibinfo{year}{2004}), \bibinfo{pages}{418--421}.
\newblock


\bibitem[Lanus et~al\mbox{.}(2021)]%
        {lanus2021combinatorial}
\bibfield{author}{\bibinfo{person}{Erin Lanus}, \bibinfo{person}{Laura~J Freeman}, \bibinfo{person}{D~Richard Kuhn}, {and} \bibinfo{person}{Raghu~N Kacker}.} \bibinfo{year}{2021}\natexlab{}.
\newblock \showarticletitle{Combinatorial testing metrics for machine learning}. In \bibinfo{booktitle}{\emph{2021 IEEE International Conference on Software Testing, Verification and Validation Workshops (ICSTW)}}. IEEE, \bibinfo{pages}{81--84}.
\newblock


\bibitem[Lewis et~al\mbox{.}(2022)]%
        {lewis2022augur}
\bibfield{author}{\bibinfo{person}{Grace~A Lewis}, \bibinfo{person}{Sebasti{\'a}n Echeverr{\'\i}a}, \bibinfo{person}{Lena Pons}, {and} \bibinfo{person}{Jeffrey Chrabaszcz}.} \bibinfo{year}{2022}\natexlab{}.
\newblock \showarticletitle{Augur: A step towards realistic drift detection in production ml systems}. In \bibinfo{booktitle}{\emph{Proceedings of the 1st Workshop on Software Engineering for Responsible AI}}. \bibinfo{pages}{37--44}.
\newblock


\bibitem[Li et~al\mbox{.}(2022)]%
        {li2022testing}
\bibfield{author}{\bibinfo{person}{Shuyue Li}, \bibinfo{person}{Jiaqi Guo}, \bibinfo{person}{Jian-Guang Lou}, \bibinfo{person}{Ming Fan}, \bibinfo{person}{Ting Liu}, {and} \bibinfo{person}{Dongmei Zhang}.} \bibinfo{year}{2022}\natexlab{}.
\newblock \showarticletitle{Testing machine learning systems in industry: an empirical study}. In \bibinfo{booktitle}{\emph{Proceedings of the 44th International Conference on Software Engineering: Software Engineering in Practice}}. \bibinfo{pages}{263--272}.
\newblock


\bibitem[Li et~al\mbox{.}(2020)]%
        {li2020ontology}
\bibfield{author}{\bibinfo{person}{Yihao Li}, \bibinfo{person}{Jianbo Tao}, {and} \bibinfo{person}{Franz Wotawa}.} \bibinfo{year}{2020}\natexlab{}.
\newblock \showarticletitle{Ontology-based test generation for automated and autonomous driving functions}.
\newblock \bibinfo{journal}{\emph{Information and software technology}}  \bibinfo{volume}{117} (\bibinfo{year}{2020}), \bibinfo{pages}{106200}.
\newblock


\bibitem[Li et~al\mbox{.}(2019)]%
        {li2019structural}
\bibfield{author}{\bibinfo{person}{Zenan Li}, \bibinfo{person}{Xiaoxing Ma}, \bibinfo{person}{Chang Xu}, {and} \bibinfo{person}{Chun Cao}.} \bibinfo{year}{2019}\natexlab{}.
\newblock \showarticletitle{Structural coverage criteria for neural networks could be misleading}. In \bibinfo{booktitle}{\emph{2019 IEEE/ACM 41st International Conference on Software Engineering: New Ideas and Emerging Results (ICSE-NIER)}}. IEEE, \bibinfo{pages}{89--92}.
\newblock


\bibitem[Lu et~al\mbox{.}(2018)]%
        {lu2018learning}
\bibfield{author}{\bibinfo{person}{Jie Lu}, \bibinfo{person}{Anjin Liu}, \bibinfo{person}{Fan Dong}, \bibinfo{person}{Feng Gu}, \bibinfo{person}{Joao Gama}, {and} \bibinfo{person}{Guangquan Zhang}.} \bibinfo{year}{2018}\natexlab{}.
\newblock \showarticletitle{Learning under concept drift: A review}.
\newblock \bibinfo{journal}{\emph{IEEE transactions on knowledge and data engineering}} \bibinfo{volume}{31}, \bibinfo{number}{12} (\bibinfo{year}{2018}), \bibinfo{pages}{2346--2363}.
\newblock


\bibitem[Ma et~al\mbox{.}(2019)]%
        {ma2019deepct}
\bibfield{author}{\bibinfo{person}{Lei Ma}, \bibinfo{person}{Felix Juefei-Xu}, \bibinfo{person}{Minhui Xue}, \bibinfo{person}{Bo Li}, \bibinfo{person}{Li Li}, \bibinfo{person}{Yang Liu}, {and} \bibinfo{person}{Jianjun Zhao}.} \bibinfo{year}{2019}\natexlab{}.
\newblock \showarticletitle{Deepct: Tomographic combinatorial testing for deep learning systems}. In \bibinfo{booktitle}{\emph{2019 IEEE 26th International Conference on Software Analysis, Evolution and Reengineering (SANER)}}. IEEE, \bibinfo{pages}{614--618}.
\newblock


\bibitem[Ma et~al\mbox{.}(2018)]%
        {ma2018deepgauge}
\bibfield{author}{\bibinfo{person}{Lei Ma}, \bibinfo{person}{Felix Juefei-Xu}, \bibinfo{person}{Fuyuan Zhang}, \bibinfo{person}{Jiyuan Sun}, \bibinfo{person}{Minhui Xue}, \bibinfo{person}{Bo Li}, \bibinfo{person}{Chunyang Chen}, \bibinfo{person}{Ting Su}, \bibinfo{person}{Li Li}, \bibinfo{person}{Yang Liu}, {et~al\mbox{.}}} \bibinfo{year}{2018}\natexlab{}.
\newblock \showarticletitle{Deepgauge: Multi-granularity testing criteria for deep learning systems}. In \bibinfo{booktitle}{\emph{Proceedings of the 33rd ACM/IEEE international conference on automated software engineering}}. \bibinfo{pages}{120--131}.
\newblock


\bibitem[Mallick et~al\mbox{.}(2022)]%
        {mallick2022matchmaker}
\bibfield{author}{\bibinfo{person}{Ankur Mallick}, \bibinfo{person}{Kevin Hsieh}, \bibinfo{person}{Behnaz Arzani}, {and} \bibinfo{person}{Gauri Joshi}.} \bibinfo{year}{2022}\natexlab{}.
\newblock \showarticletitle{Matchmaker: Data drift mitigation in machine learning for large-scale systems}.
\newblock \bibinfo{journal}{\emph{Proceedings of Machine Learning and Systems}}  \bibinfo{volume}{4} (\bibinfo{year}{2022}), \bibinfo{pages}{77--94}.
\newblock


\bibitem[McKeeman(1998)]%
        {mckeeman1998differential}
\bibfield{author}{\bibinfo{person}{William~M McKeeman}.} \bibinfo{year}{1998}\natexlab{}.
\newblock \showarticletitle{Differential testing for software}.
\newblock \bibinfo{journal}{\emph{Digital Technical Journal}} \bibinfo{volume}{10}, \bibinfo{number}{1} (\bibinfo{year}{1998}), \bibinfo{pages}{100--107}.
\newblock


\bibitem[Meng et~al\mbox{.}(2016)]%
        {meng2016mllib}
\bibfield{author}{\bibinfo{person}{Xiangrui Meng}, \bibinfo{person}{Joseph Bradley}, \bibinfo{person}{Burak Yavuz}, \bibinfo{person}{Evan Sparks}, \bibinfo{person}{Shivaram Venkataraman}, \bibinfo{person}{Davies Liu}, \bibinfo{person}{Jeremy Freeman}, \bibinfo{person}{DB Tsai}, \bibinfo{person}{Manish Amde}, \bibinfo{person}{Sean Owen}, {et~al\mbox{.}}} \bibinfo{year}{2016}\natexlab{}.
\newblock \showarticletitle{Mllib: Machine learning in apache spark}.
\newblock \bibinfo{journal}{\emph{The journal of machine learning research}} \bibinfo{volume}{17}, \bibinfo{number}{1} (\bibinfo{year}{2016}), \bibinfo{pages}{1235--1241}.
\newblock


\bibitem[Murphy et~al\mbox{.}(2008)]%
        {murphy2008properties}
\bibfield{author}{\bibinfo{person}{Christian Murphy}, \bibinfo{person}{Gail~E Kaiser}, {and} \bibinfo{person}{Lifeng Hu}.} \bibinfo{year}{2008}\natexlab{}.
\newblock \showarticletitle{Properties of machine learning applications for use in metamorphic testing}.
\newblock  (\bibinfo{year}{2008}).
\newblock


\bibitem[Nahar et~al\mbox{.}(2023)]%
        {nahar2023meta}
\bibfield{author}{\bibinfo{person}{Nadia Nahar}, \bibinfo{person}{Haoran Zhang}, \bibinfo{person}{Grace Lewis}, \bibinfo{person}{Shurui Zhou}, {and} \bibinfo{person}{Christian K{\"a}stner}.} \bibinfo{year}{2023}\natexlab{}.
\newblock \showarticletitle{A Meta-Summary of Challenges in Building Products with ML Components--Collecting Experiences from 4758+ Practitioners}.
\newblock \bibinfo{journal}{\emph{arXiv preprint arXiv:2304.00078}} (\bibinfo{year}{2023}).
\newblock


\bibitem[Nahar et~al\mbox{.}(2022)]%
        {nahar2022collaboration}
\bibfield{author}{\bibinfo{person}{Nadia Nahar}, \bibinfo{person}{Shurui Zhou}, \bibinfo{person}{Grace Lewis}, {and} \bibinfo{person}{Christian K{\"a}stner}.} \bibinfo{year}{2022}\natexlab{}.
\newblock \showarticletitle{Collaboration challenges in building ml-enabled systems: Communication, documentation, engineering, and process}. In \bibinfo{booktitle}{\emph{Proceedings of the 44th International Conference on Software Engineering}}. \bibinfo{pages}{413--425}.
\newblock


\bibitem[Nelson et~al\mbox{.}(2015)]%
        {nelson2015evaluating}
\bibfield{author}{\bibinfo{person}{Kevin Nelson}, \bibinfo{person}{George Corbin}, \bibinfo{person}{Mark Anania}, \bibinfo{person}{Matthew Kovacs}, \bibinfo{person}{Jeremy Tobias}, {and} \bibinfo{person}{Misty Blowers}.} \bibinfo{year}{2015}\natexlab{}.
\newblock \showarticletitle{Evaluating model drift in machine learning algorithms}. In \bibinfo{booktitle}{\emph{2015 IEEE Symposium on Computational Intelligence for Security and Defense Applications (CISDA)}}. IEEE, \bibinfo{pages}{1--8}.
\newblock


\bibitem[Odena et~al\mbox{.}(2019)]%
        {odena2019tensorfuzz}
\bibfield{author}{\bibinfo{person}{Augustus Odena}, \bibinfo{person}{Catherine Olsson}, \bibinfo{person}{David Andersen}, {and} \bibinfo{person}{Ian Goodfellow}.} \bibinfo{year}{2019}\natexlab{}.
\newblock \showarticletitle{Tensorfuzz: Debugging neural networks with coverage-guided fuzzing}. In \bibinfo{booktitle}{\emph{International Conference on Machine Learning}}. PMLR, \bibinfo{pages}{4901--4911}.
\newblock


\bibitem[Openja et~al\mbox{.}(2022)]%
        {openja2022empirical}
\bibfield{author}{\bibinfo{person}{Moses Openja}, \bibinfo{person}{Amin Nikanjam}, \bibinfo{person}{Ahmed~Haj Yahmed}, \bibinfo{person}{Foutse Khomh}, {and} \bibinfo{person}{Zhen Ming~Jack Jiang}.} \bibinfo{year}{2022}\natexlab{}.
\newblock \showarticletitle{An empirical study of challenges in converting deep learning models}. In \bibinfo{booktitle}{\emph{2022 IEEE International Conference on Software Maintenance and Evolution (ICSME)}}. IEEE, \bibinfo{pages}{13--23}.
\newblock


\bibitem[Patel et~al\mbox{.}(2022)]%
        {patel2022combinatorial}
\bibfield{author}{\bibinfo{person}{Ankita~Ramjibhai Patel}, \bibinfo{person}{Jaganmohan Chandrasekaran}, \bibinfo{person}{Yu Lei}, \bibinfo{person}{Raghu~N Kacker}, {and} \bibinfo{person}{D~Richard Kuhn}.} \bibinfo{year}{2022}\natexlab{}.
\newblock \showarticletitle{A combinatorial approach to fairness testing of machine learning models}. In \bibinfo{booktitle}{\emph{2022 IEEE International Conference on Software Testing, Verification and Validation Workshops (ICSTW)}}. IEEE, \bibinfo{pages}{94--101}.
\newblock


\bibitem[Pedregosa et~al\mbox{.}(2011)]%
        {scikit-learn}
\bibfield{author}{\bibinfo{person}{F. Pedregosa}, \bibinfo{person}{G. Varoquaux}, \bibinfo{person}{A. Gramfort}, \bibinfo{person}{V. Michel}, \bibinfo{person}{B. Thirion}, \bibinfo{person}{O. Grisel}, \bibinfo{person}{M. Blondel}, \bibinfo{person}{P. Prettenhofer}, \bibinfo{person}{R. Weiss}, \bibinfo{person}{V. Dubourg}, \bibinfo{person}{J. Vanderplas}, \bibinfo{person}{A. Passos}, \bibinfo{person}{D. Cournapeau}, \bibinfo{person}{M. Brucher}, \bibinfo{person}{M. Perrot}, {and} \bibinfo{person}{E. Duchesnay}.} \bibinfo{year}{2011}\natexlab{}.
\newblock \showarticletitle{Scikit-learn: Machine Learning in {P}ython}.
\newblock \bibinfo{journal}{\emph{Journal of Machine Learning Research}}  \bibinfo{volume}{12} (\bibinfo{year}{2011}), \bibinfo{pages}{2825--2830}.
\newblock


\bibitem[Pei et~al\mbox{.}(2017)]%
        {pei2017deepxplore}
\bibfield{author}{\bibinfo{person}{Kexin Pei}, \bibinfo{person}{Yinzhi Cao}, \bibinfo{person}{Junfeng Yang}, {and} \bibinfo{person}{Suman Jana}.} \bibinfo{year}{2017}\natexlab{}.
\newblock \showarticletitle{Deepxplore: Automated whitebox testing of deep learning systems}. In \bibinfo{booktitle}{\emph{proceedings of the 26th Symposium on Operating Systems Principles}}. \bibinfo{pages}{1--18}.
\newblock


\bibitem[Pham et~al\mbox{.}(2019)]%
        {pham2019cradle}
\bibfield{author}{\bibinfo{person}{Hung~Viet Pham}, \bibinfo{person}{Thibaud Lutellier}, \bibinfo{person}{Weizhen Qi}, {and} \bibinfo{person}{Lin Tan}.} \bibinfo{year}{2019}\natexlab{}.
\newblock \showarticletitle{CRADLE: cross-backend validation to detect and localize bugs in deep learning libraries}. In \bibinfo{booktitle}{\emph{2019 IEEE/ACM 41st International Conference on Software Engineering (ICSE)}}. IEEE, \bibinfo{pages}{1027--1038}.
\newblock


\bibitem[Poenaru-Olaru et~al\mbox{.}(2023)]%
        {poenaru2023maintaining}
\bibfield{author}{\bibinfo{person}{Lorena Poenaru-Olaru}, \bibinfo{person}{Luis Cruz}, \bibinfo{person}{Jan~S Rellermeyer}, {and} \bibinfo{person}{Arie Van~Deursen}.} \bibinfo{year}{2023}\natexlab{}.
\newblock \showarticletitle{Maintaining and Monitoring AIOps Models Against Concept Drift}. In \bibinfo{booktitle}{\emph{2023 IEEE/ACM 2nd International Conference on AI Engineering--Software Engineering for AI (CAIN)}}. IEEE, \bibinfo{pages}{98--99}.
\newblock


\bibitem[Poenaru-Olaru et~al\mbox{.}(2022)]%
        {poenaru2022concept}
\bibfield{author}{\bibinfo{person}{Lorena Poenaru-Olaru}, \bibinfo{person}{Luis Cruz}, \bibinfo{person}{Arie van Deursen}, {and} \bibinfo{person}{Jan~S Rellermeyer}.} \bibinfo{year}{2022}\natexlab{}.
\newblock \showarticletitle{Are concept drift detectors reliable alarming systems?-a comparative study}. In \bibinfo{booktitle}{\emph{2022 IEEE International Conference on Big Data (Big Data)}}. IEEE, \bibinfo{pages}{3364--3373}.
\newblock


\bibitem[Ramanagopal et~al\mbox{.}(2018)]%
        {ramanagopal2018failing}
\bibfield{author}{\bibinfo{person}{Manikandasriram~Srinivasan Ramanagopal}, \bibinfo{person}{Cyrus Anderson}, \bibinfo{person}{Ram Vasudevan}, {and} \bibinfo{person}{Matthew Johnson-Roberson}.} \bibinfo{year}{2018}\natexlab{}.
\newblock \showarticletitle{Failing to learn: Autonomously identifying perception failures for self-driving cars}.
\newblock \bibinfo{journal}{\emph{IEEE Robotics and Automation Letters}} \bibinfo{volume}{3}, \bibinfo{number}{4} (\bibinfo{year}{2018}), \bibinfo{pages}{3860--3867}.
\newblock


\bibitem[Riccio et~al\mbox{.}(2020)]%
        {riccio2020testing}
\bibfield{author}{\bibinfo{person}{Vincenzo Riccio}, \bibinfo{person}{Gunel Jahangirova}, \bibinfo{person}{Andrea Stocco}, \bibinfo{person}{Nargiz Humbatova}, \bibinfo{person}{Michael Weiss}, {and} \bibinfo{person}{Paolo Tonella}.} \bibinfo{year}{2020}\natexlab{}.
\newblock \showarticletitle{Testing machine learning based systems: a systematic mapping}.
\newblock \bibinfo{journal}{\emph{Empirical Software Engineering}}  \bibinfo{volume}{25} (\bibinfo{year}{2020}), \bibinfo{pages}{5193--5254}.
\newblock


\bibitem[Segura et~al\mbox{.}(2018)]%
        {segura2018metamorphic}
\bibfield{author}{\bibinfo{person}{Sergio Segura}, \bibinfo{person}{Dave Towey}, \bibinfo{person}{Zhi~Quan Zhou}, {and} \bibinfo{person}{Tsong~Yueh Chen}.} \bibinfo{year}{2018}\natexlab{}.
\newblock \showarticletitle{Metamorphic testing: Testing the untestable}.
\newblock \bibinfo{journal}{\emph{IEEE Software}} \bibinfo{volume}{37}, \bibinfo{number}{3} (\bibinfo{year}{2018}), \bibinfo{pages}{46--53}.
\newblock


\bibitem[Shivashankar and Martini(2022)]%
        {shivashankar2022maintainability}
\bibfield{author}{\bibinfo{person}{Karthik Shivashankar} {and} \bibinfo{person}{Antonio Martini}.} \bibinfo{year}{2022}\natexlab{}.
\newblock \showarticletitle{Maintainability {C}hallenges in {ML}: A {S}ystematic {L}iterature {R}eview}. In \bibinfo{booktitle}{\emph{2022 48th Euromicro Conference on Software Engineering and Advanced Applications (SEAA)}}. IEEE, \bibinfo{pages}{60--67}.
\newblock


\bibitem[Szegedy et~al\mbox{.}(2013)]%
        {szegedy2013intriguing}
\bibfield{author}{\bibinfo{person}{Christian Szegedy}, \bibinfo{person}{Wojciech Zaremba}, \bibinfo{person}{Ilya Sutskever}, \bibinfo{person}{Joan Bruna}, \bibinfo{person}{Dumitru Erhan}, \bibinfo{person}{Ian Goodfellow}, {and} \bibinfo{person}{Rob Fergus}.} \bibinfo{year}{2013}\natexlab{}.
\newblock \showarticletitle{Intriguing properties of neural networks}.
\newblock \bibinfo{journal}{\emph{arXiv preprint arXiv:1312.6199}} (\bibinfo{year}{2013}).
\newblock


\bibitem[Tian et~al\mbox{.}(2018)]%
        {tian2018deeptest}
\bibfield{author}{\bibinfo{person}{Yuchi Tian}, \bibinfo{person}{Kexin Pei}, \bibinfo{person}{Suman Jana}, {and} \bibinfo{person}{Baishakhi Ray}.} \bibinfo{year}{2018}\natexlab{}.
\newblock \showarticletitle{Deeptest: Automated testing of deep-neural-network-driven autonomous cars}. In \bibinfo{booktitle}{\emph{Proceedings of the 40th international conference on software engineering}}. \bibinfo{pages}{303--314}.
\newblock


\bibitem[Tuncali et~al\mbox{.}(2018)]%
        {tuncali2018simulation}
\bibfield{author}{\bibinfo{person}{Cumhur~Erkan Tuncali}, \bibinfo{person}{Georgios Fainekos}, \bibinfo{person}{Hisahiro Ito}, {and} \bibinfo{person}{James Kapinski}.} \bibinfo{year}{2018}\natexlab{}.
\newblock \showarticletitle{Simulation-based adversarial test generation for autonomous vehicles with machine learning components}. In \bibinfo{booktitle}{\emph{2018 IEEE Intelligent Vehicles Symposium (IV)}}. IEEE, \bibinfo{pages}{1555--1562}.
\newblock


\bibitem[Wan et~al\mbox{.}(2019)]%
        {wan2019does}
\bibfield{author}{\bibinfo{person}{Zhiyuan Wan}, \bibinfo{person}{Xin Xia}, \bibinfo{person}{David Lo}, {and} \bibinfo{person}{Gail~C Murphy}.} \bibinfo{year}{2019}\natexlab{}.
\newblock \showarticletitle{How does machine learning change software development practices?}
\newblock \bibinfo{journal}{\emph{IEEE Transactions on Software Engineering}} \bibinfo{volume}{47}, \bibinfo{number}{9} (\bibinfo{year}{2019}), \bibinfo{pages}{1857--1871}.
\newblock


\bibitem[Webb et~al\mbox{.}(2016)]%
        {webb2016characterizing}
\bibfield{author}{\bibinfo{person}{Geoffrey~I Webb}, \bibinfo{person}{Roy Hyde}, \bibinfo{person}{Hong Cao}, \bibinfo{person}{Hai~Long Nguyen}, {and} \bibinfo{person}{Francois Petitjean}.} \bibinfo{year}{2016}\natexlab{}.
\newblock \showarticletitle{Characterizing concept drift}.
\newblock \bibinfo{journal}{\emph{Data Mining and Knowledge Discovery}} \bibinfo{volume}{30}, \bibinfo{number}{4} (\bibinfo{year}{2016}), \bibinfo{pages}{964--994}.
\newblock


\bibitem[Wei et~al\mbox{.}(2022)]%
        {wei2022free}
\bibfield{author}{\bibinfo{person}{Anjiang Wei}, \bibinfo{person}{Yinlin Deng}, \bibinfo{person}{Chenyuan Yang}, {and} \bibinfo{person}{Lingming Zhang}.} \bibinfo{year}{2022}\natexlab{}.
\newblock \showarticletitle{Free lunch for testing: Fuzzing deep-learning libraries from open source}. In \bibinfo{booktitle}{\emph{Proceedings of the 44th International Conference on Software Engineering}}. \bibinfo{pages}{995--1007}.
\newblock


\bibitem[Xie et~al\mbox{.}(2022)]%
        {xie2022docter}
\bibfield{author}{\bibinfo{person}{Danning Xie}, \bibinfo{person}{Yitong Li}, \bibinfo{person}{Mijung Kim}, \bibinfo{person}{Hung~Viet Pham}, \bibinfo{person}{Lin Tan}, \bibinfo{person}{Xiangyu Zhang}, {and} \bibinfo{person}{Michael~W Godfrey}.} \bibinfo{year}{2022}\natexlab{}.
\newblock \showarticletitle{DocTer: documentation-guided fuzzing for testing deep learning API functions}. In \bibinfo{booktitle}{\emph{Proceedings of the 31st ACM SIGSOFT International Symposium on Software Testing and Analysis}}. \bibinfo{pages}{176--188}.
\newblock


\bibitem[Xie et~al\mbox{.}(2019)]%
        {xie2019deephunter}
\bibfield{author}{\bibinfo{person}{Xiaofei Xie}, \bibinfo{person}{Lei Ma}, \bibinfo{person}{Felix Juefei-Xu}, \bibinfo{person}{Minhui Xue}, \bibinfo{person}{Hongxu Chen}, \bibinfo{person}{Yang Liu}, \bibinfo{person}{Jianjun Zhao}, \bibinfo{person}{Bo Li}, \bibinfo{person}{Jianxiong Yin}, {and} \bibinfo{person}{Simon See}.} \bibinfo{year}{2019}\natexlab{}.
\newblock \showarticletitle{Deephunter: a coverage-guided fuzz testing framework for deep neural networks}. In \bibinfo{booktitle}{\emph{Proceedings of the 28th ACM SIGSOFT International Symposium on Software Testing and Analysis}}. \bibinfo{pages}{146--157}.
\newblock


\bibitem[Xie et~al\mbox{.}(2020)]%
        {xie2020mettle}
\bibfield{author}{\bibinfo{person}{Xiaoyuan Xie}, \bibinfo{person}{Zhiyi Zhang}, \bibinfo{person}{Tsong~Yueh Chen}, \bibinfo{person}{Yang Liu}, \bibinfo{person}{Pak-Lok Poon}, {and} \bibinfo{person}{Baowen Xu}.} \bibinfo{year}{2020}\natexlab{}.
\newblock \showarticletitle{METTLE: a METamorphic testing approach to assessing and validating unsupervised machine LEarning systems}.
\newblock \bibinfo{journal}{\emph{IEEE Transactions on Reliability}} \bibinfo{volume}{69}, \bibinfo{number}{4} (\bibinfo{year}{2020}), \bibinfo{pages}{1293--1322}.
\newblock


\bibitem[Yang et~al\mbox{.}(2023)]%
        {yang2023fuzzing}
\bibfield{author}{\bibinfo{person}{Chenyuan Yang}, \bibinfo{person}{Yinlin Deng}, \bibinfo{person}{Jiayi Yao}, \bibinfo{person}{Yuxing Tu}, \bibinfo{person}{Hanchi Li}, {and} \bibinfo{person}{Lingming Zhang}.} \bibinfo{year}{2023}\natexlab{}.
\newblock \showarticletitle{Fuzzing automatic differentiation in deep-learning libraries}.
\newblock \bibinfo{journal}{\emph{arXiv preprint arXiv:2302.04351}} (\bibinfo{year}{2023}).
\newblock


\bibitem[Yang et~al\mbox{.}(2021)]%
        {yang2021cade}
\bibfield{author}{\bibinfo{person}{Limin Yang}, \bibinfo{person}{Wenbo Guo}, \bibinfo{person}{Qingying Hao}, \bibinfo{person}{Arridhana Ciptadi}, \bibinfo{person}{Ali Ahmadzadeh}, \bibinfo{person}{Xinyu Xing}, {and} \bibinfo{person}{Gang Wang}.} \bibinfo{year}{2021}\natexlab{}.
\newblock \showarticletitle{$\{$CADE$\}$: Detecting and explaining concept drift samples for security applications}. In \bibinfo{booktitle}{\emph{30th USENIX Security Symposium (USENIX Security 21)}}. \bibinfo{pages}{2327--2344}.
\newblock


\bibitem[Zhang et~al\mbox{.}(2020)]%
        {zhang2020machine}
\bibfield{author}{\bibinfo{person}{Jie~M Zhang}, \bibinfo{person}{Mark Harman}, \bibinfo{person}{Lei Ma}, {and} \bibinfo{person}{Yang Liu}.} \bibinfo{year}{2020}\natexlab{}.
\newblock \showarticletitle{Machine learning testing: Survey, landscapes and horizons}.
\newblock \bibinfo{journal}{\emph{IEEE Transactions on Software Engineering}} \bibinfo{volume}{48}, \bibinfo{number}{1} (\bibinfo{year}{2020}), \bibinfo{pages}{1--36}.
\newblock


\bibitem[Zhang et~al\mbox{.}(2018)]%
        {zhang2018deeproad}
\bibfield{author}{\bibinfo{person}{Mengshi Zhang}, \bibinfo{person}{Yuqun Zhang}, \bibinfo{person}{Lingming Zhang}, \bibinfo{person}{Cong Liu}, {and} \bibinfo{person}{Sarfraz Khurshid}.} \bibinfo{year}{2018}\natexlab{}.
\newblock \showarticletitle{Deep{R}oad: {GAN}-based metamorphic testing and input validation framework for autonomous driving systems}. In \bibinfo{booktitle}{\emph{Proceedings of the 33rd ACM/IEEE International Conference on Automated Software Engineering}}. \bibinfo{pages}{132--142}.
\newblock


\bibitem[Zhang et~al\mbox{.}(2019)]%
        {zhang2019software}
\bibfield{author}{\bibinfo{person}{Xufan Zhang}, \bibinfo{person}{Yilin Yang}, \bibinfo{person}{Yang Feng}, {and} \bibinfo{person}{Zhenyu Chen}.} \bibinfo{year}{2019}\natexlab{}.
\newblock \showarticletitle{Software {E}ngineering {P}ractice in the {D}evelopment of {D}eep {L}earning {A}pplications}.
\newblock \bibinfo{journal}{\emph{arXiv preprint arXiv:1910.03156}} (\bibinfo{year}{2019}).
\newblock


\bibitem[Zhengxin et~al\mbox{.}(2023)]%
        {zhengxin2023mlops}
\bibfield{author}{\bibinfo{person}{Fang Zhengxin}, \bibinfo{person}{Yuan Yi}, \bibinfo{person}{Zhang Jingyu}, \bibinfo{person}{Liu Yue}, \bibinfo{person}{Mu Yuechen}, \bibinfo{person}{Lu Qinghua}, \bibinfo{person}{Xu Xiwei}, \bibinfo{person}{Wang Jeff}, \bibinfo{person}{Wang Chen}, \bibinfo{person}{Zhang Shuai}, {et~al\mbox{.}}} \bibinfo{year}{2023}\natexlab{}.
\newblock \showarticletitle{MLOps Spanning Whole Machine Learning Life Cycle: A Survey}.
\newblock \bibinfo{journal}{\emph{arXiv preprint arXiv:2304.07296}} (\bibinfo{year}{2023}).
\newblock


\bibitem[Zhou et~al\mbox{.}(2020)]%
        {zhou2020deepbillboard}
\bibfield{author}{\bibinfo{person}{Husheng Zhou}, \bibinfo{person}{Wei Li}, \bibinfo{person}{Zelun Kong}, \bibinfo{person}{Junfeng Guo}, \bibinfo{person}{Yuqun Zhang}, \bibinfo{person}{Bei Yu}, \bibinfo{person}{Lingming Zhang}, {and} \bibinfo{person}{Cong Liu}.} \bibinfo{year}{2020}\natexlab{}.
\newblock \showarticletitle{Deepbillboard: Systematic physical-world testing of autonomous driving systems}. In \bibinfo{booktitle}{\emph{Proceedings of the ACM/IEEE 42nd International Conference on Software Engineering}}. \bibinfo{pages}{347--358}.
\newblock


\bibitem[Zhou and Sun(2019)]%
        {zhou2019metamorphic}
\bibfield{author}{\bibinfo{person}{Zhi~Quan Zhou} {and} \bibinfo{person}{Liqun Sun}.} \bibinfo{year}{2019}\natexlab{}.
\newblock \showarticletitle{Metamorphic testing of driverless cars}.
\newblock \bibinfo{journal}{\emph{Commun. ACM}} \bibinfo{volume}{62}, \bibinfo{number}{3} (\bibinfo{year}{2019}), \bibinfo{pages}{61--67}.
\newblock


\bibitem[Zhu et~al\mbox{.}(2020)]%
        {zhu2020more}
\bibfield{author}{\bibinfo{person}{Tianqing Zhu}, \bibinfo{person}{Dayong Ye}, \bibinfo{person}{Wei Wang}, \bibinfo{person}{Wanlei Zhou}, {and} \bibinfo{person}{S~Yu Philip}.} \bibinfo{year}{2020}\natexlab{}.
\newblock \showarticletitle{More than privacy: Applying differential privacy in key areas of artificial intelligence}.
\newblock \bibinfo{journal}{\emph{IEEE Transactions on Knowledge and Data Engineering}} \bibinfo{volume}{34}, \bibinfo{number}{6} (\bibinfo{year}{2020}), \bibinfo{pages}{2824--2843}.
\newblock


\bibitem[{\v{Z}}liobait{\.e} et~al\mbox{.}(2016)]%
        {vzliobaite2016overview}
\bibfield{author}{\bibinfo{person}{Indr{\.e} {\v{Z}}liobait{\.e}}, \bibinfo{person}{Mykola Pechenizkiy}, {and} \bibinfo{person}{Joao Gama}.} \bibinfo{year}{2016}\natexlab{}.
\newblock \showarticletitle{An overview of concept drift applications}.
\newblock \bibinfo{journal}{\emph{Big data analysis: new algorithms for a new society}} (\bibinfo{year}{2016}), \bibinfo{pages}{91--114}.
\newblock


\end{thebibliography}

\end{document}